\documentclass[10pt,letterpaper]{article}
\usepackage{opex3}
\usepackage[utf8x]{inputenc}
\usepackage{amsmath,amssymb}
\usepackage{graphicx}
\usepackage{subfigure}
\usepackage{enumerate}
\usepackage{cite}

\begin{document}

\title{Improved success rate and stability for phase retrieval by including randomized overrelaxation in the hybrid input output algorithm}

\author{Martin K\"{o}hl,\textsuperscript{*} A. A. Minkevich, and Tilo Baumbach}

\address{Institut f\"{u}r Synchrotronstrahlung, Karlsruher Institut f\"{u}r Technologie, Hermann-von-Helmholtz-Platz 1, D-76344 Eggenstein-Leopoldshafen, Germany}

\email{\textsuperscript{*}martin.koehl@kit.edu} 

\newcommand{\ER}{ER}
\newcommand{\HIO}{HIO}
\newcommand{\HIOO}{HIO+O}
\newcommand{\HIOOR}{HIO+$\text{O}_{\text{R}}$}
\newcommand{\HIOER}{HIO+ER}
\newcommand{\HIOERO}{(HIO+O)+ER}
\newcommand{\HIOEROR}{(HIO+$\text{O}_{\text{R}}$)+ER}
\newcommand{\vecx}{\vec{x}}
\newcommand{\veck}{\vec{k}}
\newcommand{\imagUnit}{\mathrm{i}}
\newcommand{\Heaviside}{\Theta}
\newcommand{\ShapeRealspace}{\mathbb{S}}
\newcommand{\ShapeFunction}{\Omega}
\newcommand{\DiscretizedRegionRealspace}{\mathbb{D}}
\newcommand{\braket}[2]{\left\langle \, #1 \, ; \, #2 \, \right\rangle}
\newcommand{\fInput}{f_{\mathrm{in}}}
\newcommand{\fReconstructed}{f}
\newcommand{\gReconstructed}{\widetilde{g}}
\newcommand{\gAmp}{A}
\newcommand{\IFT}{\mathbf{IFT}}
\newcommand{\FT}{\mathbf{FT}}
\newcommand{\Project}[1]{\mathbf{P}_{#1}}
\newcommand{\OverProject}[2]{\mathbf{Q}_{#1;#2}}
\newcommand{\Identity}{1}
\newcommand{\NReal}{N_{\mathrm{Real}}}
\newcommand{\mesh}{\mathbb{M}}
\newcommand{\EROperator}{\hat{H}_{\mathrm{ER}}}
\newcommand{\HIOOperator}{\hat{H}_{\mathrm{HIO}}}
\newcommand{\HIOOOperator}{\hat{H}_{\mathrm{HIO+O}_R}}
\newcommand{\HIOROperator}{\hat{H}_{\mathrm{HIO}_R}}
\newcommand{\HProj}{\hat{H}_{\mathrm{Proj}}}
\newcommand{\DifferenceOperator}{\hat{H}_{\mathrm{Diff}}}


\begin{abstract}
In this paper, we study the success rate of the reconstruction of objects of finite extent given the magnitude of its Fourier transform and its geometrical shape. We demonstrate that the commonly used combination of the hybrid input output and error reduction algorithm is significantly outperformed by an extension of this algorithm based on randomized overrelaxation. In most cases, this extension tremendously enhances the success rate of reconstructions for a fixed number of iterations as compared to reconstructions solely based on the traditional algorithm. The good scaling properties in terms of computational time and memory requirements of the original algorithm are not influenced by this extension.
\end{abstract}




Inverse problems play an important role in many areas of physics. In an inverse problem we intend to reconstruct physical properties (``scattering potential'') of some object given the result of the interaction (``scattering'') with a well-defined input signal. Typically, propagating wave fields (e.g., acoustic waves, electromagnetic waves, \ldots) are used to scan the object. However, for rapidly oscillating electromagnetic wave fields like x-rays, detection devices are only capable of recording the average intensity of the scattered wave field. Its phase information, which also contains important information about the scattering potential, is lost during the measurement process. Despite the lost phase information, x-rays are an important, \emph{non-destructive} tool for the investigation of structural properties of matter because they exhibit high penetration depth and high spatial resolution \cite{Veen2004,Millane1990,Pietsch2004}. Note, that $3^{\text{rd}}$ generation synchrotrons \cite{Afanasiev2004} provide brilliant x-ray beams with a very high degree of coherence and with very high flux, so that even dot- and wire-like nanostructures can be investigated \cite{Millane1990,Robinson2009,Miao1999,Pfeifer2006,Biermanns2009,Minkevich07,Minkevich11EPL,Minkevich11PRB}.

In this paper, we introduce a new algorithm -- the \HIOEROR{}-algorithm -- which is capable of reconstructing the missing phase information in a variety of cases for which the traditionally and commonly used \HIOER{}-algorithm \cite{Fienup1982,Fienup1986a} (reviewed in sec. \ref{sec:retrievalAlgorithm}) fails. Before we introduce this new algorithm in sec. \ref{sec:retrievalAlgorithm}, we shortly review the phase retrieval problem in sec. \ref{sec:PhaseRetrieval}. In particular, we want to stress the fact that the \HIOEROR{}-algorithm is based on randomized overrelaxation \cite{Levi84,Youla1982} and, thus, does not require additional mathematical constraints \cite{PhysRevB.78.174110} compared to the \HIOER{}-algorithm. We reformulate our approach using projection polynomials which facilitate a straight forward investigation of randomization of the \HIO{}-algorithm without correlations in the coefficients enforced by overrelaxation. Section \ref{sec:Convergence} discusses numerical aspects how to measure the convergence of phase retrieval. Finally, sec. \ref{sec:numericalExamples} illustrates the advantages of the \HIOEROR{}-algorithm with respect to the traditional \HIOER{}-algorithms in detail. For this purpose, we focus on three numerical examples each with different specific properties and compare the success rate of reconstructions based on our \HIOEROR{}-algorithm and on the \HIOER{}-algorithm. Moreover, we investigate the sensitivity of our algorithm to the choice of its parameters. In this paper, we focus on the algorithm itself and its improved convergence properties for specific, well-known problems related to phase retrieval based on the \HIOER{}-algorithm. A separate publication will illustrate the improved features of the \HIOEROR{}-algorithm for the reconstruction of the displacement field in inhomogeneously strained nanocrystals. 

\section{The phase retrieval problem}
\label{sec:PhaseRetrieval}

Consider some unknown object $\fInput(\vecx)$ in $d$-dimensional direct space (``position space''). We assume that (i) the shape $\ShapeRealspace$ of $\fInput(\vecx)$ (i.e., the smallest domain in direct space for which $\fInput(\vecx) \neq 0$) and (ii) the quantity $\gAmp_{\veck} \equiv \left| \FT\{ \fInput(\vecx) \} \right|$ (i.e., the amplitudes of its Fourier transform) are known. It has been proven that this information is sufficient to reconstruct the object $\fInput(\vecx)$ \emph{if} the shape $\ShapeRealspace$ is \emph{finite} and the dimension $d$ of direct space is at least two \cite{Millane1990,Millane1996,Seldin1990,Bates1982}. Physically, the second condition is fulfilled in scattering experiments which can be described in lowest order Born-approximation (also called kinematical approximation). This is typically true for nanostructures illuminated by coherent x-ray beams.

Suppose we discretize a rectangular region $\DiscretizedRegionRealspace \subset \mathbb{R}^d$ in direct space, which fully contains our object $\fInput(\vecx)$ (i.e., $\ShapeRealspace \subset \DiscretizedRegionRealspace$), with $N_i$, $i \in \{1,2,\ldots,d\}$, equidistant points. The discrete Fourier transform (DFT) maps this set of points to a mesh $\mesh$ in reciprocal space which is related to the Fourier transform of our continuous direct space object $\fInput(\vecx)$ \cite{Auslander1989}. Hence, we have
\begin{equation}
 \fInput(\vecx) = \sum_{\veck \in \mesh} \left[ \prod_{j=1}^{d} \exp\left( \frac{2 \pi \imagUnit}{N_j} k_j  x_j \right) \right] \gAmp_{\veck} \exp( \imagUnit \Phi_{\veck}) \ \forall \ \vecx \notin \ShapeRealspace \ ,
\end{equation}
where $\vecx = (x_1,x_2,\ldots,x_d)^T$ corresponds to the equidistantly sampled points in direct space defined by $\DiscretizedRegionRealspace$ and $\Phi_{\veck}$ are the phases associated with the amplitudes $\gAmp_{\veck}$ in reciprocal space. 

In direct space, every point $\vecx$ inside the object domain $\ShapeRealspace$ corresponds to (up to) two real unknowns (magnitude and phase). However, outside the domain $\ShapeRealspace$, $\fInput(\vecx)$ is precisely zero for all positions $\vecx$. Hence, each point outside $\ShapeRealspace$ defines two real equations for the determination of the $N$ real phases $\Phi_{\veck}$, if we require $ \fInput(\vecx) = 0 \ \forall \ \vecx \notin \ShapeRealspace$. Therefore, we can hope to reconstruct the missing phases $\Phi_{\veck}$ on the sampled grid $\mesh$ once
\begin{equation}
2 (N-N_{\ShapeRealspace} ) \geq N \quad \leftrightarrow \quad N \geq 2 N_{\ShapeRealspace} \ ,
\label{eqn:RequirementDataPointsPhaseRetrieval}
\end{equation}
where $N_{\ShapeRealspace}$ is the number of points inside the region $\ShapeRealspace$. The fraction $\sigma = N / N_{\ShapeRealspace} $ is called oversampling ratio. From our elementary discussion, we can conclude that $\sigma=2$ is a lower bound for a successful reconstruction, if no additional a priori knowledge is available. For more details on oversampling, we refer the reader to the investigations of Elser et Milliane \cite{Elser2008} and of Miao et al. \cite{Miao1998} (and the references therein).

With that knowledge, we restate the phase retrieval problem more precisely: Given the shape $\ShapeRealspace$ of some object $\fInput(\vecx)$ and the modulus $\gAmp_{\veck}$ of the Fourier transform of $\fInput(\vecx)$ on a mesh $\mesh$ that satisfies Eq. \eqref{eqn:RequirementDataPointsPhaseRetrieval}, we try to reconstruct the phases $\Phi_{\veck}$ and, thus, the object $\fInput(\vecx)$. 

Note, that the reconstruction is only unique up to the unavoidable inherent symmetries of the Fourier transform \cite{Millane1990,Millane1996,Seldin1990}, i.e., a global phase shift (typically irrelevant), a plane wave modulation $e^{\imagUnit \veck \cdot \vecx}$ in reciprocal space (fixed by the position of the shape $\ShapeRealspace$ inside $\DiscretizedRegionRealspace$) and the degeneracy between the object $f_{\vecx}$ and its complex conjugated, inversion symmetric object $f_{\vecx}^{*}$. This last ambiguity constitutes a severe problem if the shape $\ShapeRealspace$ is inversion symmetric, but the object $\fInput$ does not possess this symmetry \cite{Fienup1986a,Fienup1986b}. For the remainder of this paper, we restrict to domains $\ShapeRealspace$ which are not inversion symmetric. However, even if we exclude objects with inversion symmetric shape, it is highly non-trivial to formulate a phase retrieval algorithm that works \emph{automatically} without manual supervision and tuning of parameters during the reconstruction. The \HIOEROR{}-algorithm which we propose in the next section is capable of performing this task for a large class of objects $\fInput$. 

\section{Reconstruction algorithms}
\label{sec:retrievalAlgorithm}

In this section, we will shortly review the traditional combination of the hybrid input output and error reduction (\HIOER{}) algorithm \cite{Fienup1982,Fienup1986a} and propose our extension to this algorithm. Throughout the entire reconstruction process, we assume that we know the exact geometrical shape $\ShapeRealspace$ of the object which we reconstruct.

\subsection{Hybrid input output and error reduction}

A fast and efficient iterative phase retrieval algorithm is based on alternating projections of a trial solution onto the constraints in direct space and in reciprocal space. Hence, a single iterative step $(i)$ of this algorithm, which is called error reduction (\ER{}) algorithm \cite{Fienup1982}, is defined by
\begin{equation}
 f^{(i+1)}_{\vecx} = \Project{\ShapeRealspace}  \Project{\gAmp} f^{(i)}_{\vecx} \equiv \EROperator f^{(i)}_{\vecx} \ .
\end{equation}
Typically, initial phases $\Phi_{\veck}^{(0)}$ are chosen randomly. Here, $\Project{\ShapeRealspace}$ and $\Project{\gAmp}$ are projection operators in direct space and reciprocal space respectively, i.e.,
\begin{subequations}
\begin{equation} 
 \Project{\ShapeRealspace} f_{\vecx}^{(i)} = \left\{ 
\begin{array}{ll}
 f_{\vecx}^{(i)} & \text{ if } \vecx \in \ShapeRealspace \ , \\
 0 & \text{ if } \vecx \notin \ShapeRealspace \ 
\end{array}
\right.
\end{equation}
and 
\begin{equation}
 \Project{\gAmp} g_{\veck}^{(i)} = \gAmp_{\veck} \exp\left( \imagUnit \arg \left( g_{\veck}^{(i)} \right) \right) \ .
\end{equation}
\end{subequations}
For notational simplicity, we assume that any operand of $ \Project{\gAmp}$ or $ \Project{\ShapeRealspace}$ is transformed to the proper space (i.e., Fourier transform to reciprocal or direct space respectively) before the operator $ \Project{\gAmp}$ or $ \Project{\ShapeRealspace}$ is applied (e.g., $\Project{\gAmp} f_{\vecx}^{(i)}$ means $\Project{\gAmp} \ \FT\{ f_{\vecx}^{(i)} \} $). 

The projection operator $\Project{\gAmp}$ is \emph{non-linear}, \emph{non-convex} and \emph{non-unique} \cite{Bauschke02}. The values $|g_{\veck}^{(i)}|$ are important for determining the difference to the (experimentally accessible) input data $\gAmp_{\veck}$ (see Eq. \eqref{eqn:DefErrorMetrics}). The \ER{}-algorithm is a local minimizer of a suitable chosen error metrics \cite{Fienup1982,Fienup1986a}. However, in practice, phase retrieval problems involve many local minima. Hence, the \ER{}-algorithm will in general not converge to the correct solution $\fInput(\vecx)$, but to some local minimum. In addition, the error metric may stagnate for many iterations before decreasing further. More information on stagnation problems and the \ER{}-algorithm can be found in \cite{Fienup1982,Fienup1986a,Fienup1986b}. 

Due to these problems, further algorithms, which try to avoid stagnation and convergence to local minima, have been developed. A very important algorithm is the hybrid input output (\HIO{}) algorithm proposed by Fienup \cite{Fienup1982,Fienup1986a,Fienup1986b}. This algorithm is also an iterative procedure which is defined by the map
\begin{equation}
 f^{(i+1)}_{\vecx} = \left\{ 
\begin{array}{ll}
 \Project{\gAmp} f^{(i)}_{\vecx} & \text{ if } \vecx \in \ShapeRealspace \ , \\
 f^{(i)}_{\vecx} - \beta \Project{\gAmp} f^{(i)}_{\vecx} & \text{ if } \vecx \notin \ShapeRealspace \ , 
\end{array}
\right.
\label{eqn:HIORealSpaceAssembly}
\end{equation}
where $\beta$ is a real parameter typically chosen in the range $[0.5;1.0]$ \cite{Fienup1986b}. Note, that this definition can be rewritten as
\begin{equation}
 f^{(i+1)}_{\vecx} = \left[ \Identity - \Project{\ShapeRealspace} - \beta \Project{\gAmp} + \left( 1 + \beta \right) \Project{\ShapeRealspace}  \Project{\gAmp} \right] f^{(i)}_{\vecx} \equiv \HIOOperator(\beta) f^{(i)}_{\vecx} \ .
\label{eqn:HIOPolynomial}
\end{equation}

The convergence properties of the \HIO{}-algorithm have been investigated in more detail in \cite{Takajo98,Takajo99,Miao1998}. 

This algorithm is much stronger in avoiding stagnation and local minima. However, the \HIO{}-algorithm shows its full potential only if it is combined with the \ER{}-algorithm. One step of this combined \HIOER{}-algorithm consists of $N_{\mathrm{\HIO}}$ iterations of the \HIO-algorithm followed by $N_{\mathrm{\ER}}$ iterations of the \ER{}-algorithm. This combination is more successful than both algorithms on their own as it was observed in practice \cite{Fienup1986a,Fienup1986b}. Although this algorithm is already very powerful, it is not yet satisfactory for many objects as we will show in sec. \ref{sec:numericalExamples}.

\subsection{Hybrid input output with randomized overrelaxation (\HIOEROR{})}

The \HIOER{}-algorithm can be further improved by including randomized overrelaxation \cite{Levi84,Youla1982,Elser03}. Basically, overrelaxation corresponds to replacing a projection operator $\Project{\mu}$ by 
\begin{equation}
  \OverProject{\mu}{\lambda_\mu} \equiv \Identity + \lambda_\mu \left( \Project{\mu} - \Identity \right) \ ,
\end{equation}
where $\lambda_\mu$ is a real constant called \emph{relaxation parameter} \cite{Levi84}. The limiting case $ \OverProject{\mu}{\lambda_\mu} \rightarrow \Project{\mu}$ is obtained for $ \lambda_\mu \rightarrow 1$.

Overrelaxation (without any randomization) has been investigated for convex problems \cite{Youla1982} and in connection with the \ER{}-algorithm for phase retrieval \cite{Levi84}. Moreover, overrelaxation is also included in the difference map algorithm proposed by Elser in \cite{Elser03}. In the difference map algorithm with overrelaxation, the iterative step for our set of constraints is given by \cite{Marchesini2007Rev}
\begin{equation}
 f^{(i+1)}_{\vecx} = \left[ \Identity + \beta \left( \Project{\ShapeRealspace} \OverProject{\gAmp}{\lambda_{\gAmp}} - \Project{\gAmp} \OverProject{\ShapeRealspace}{\lambda_{\ShapeRealspace}} \right) \right] f^{(i)}_{\vecx} \equiv \DifferenceOperator(\beta,\lambda_{\gAmp},\lambda_{\ShapeRealspace}) f^{(i)}_{\vecx} \ ,
\label{eqn:DifferenceMapPolynomial}
\end{equation}
where Elser proposes to choose the parameters as $\lambda_{\gAmp} = \lambda_{\ShapeRealspace} = \beta^{-1}$ \cite{Elser03}. A very nice comparison of several phase retrieval algorithms and a benchmark thereof can be found in \cite{Marchesini2007Rev}. Marchesini demonstrated in \cite{Marchesini07} that for his particular numerical example an additional low-dimensional subspace saddle-point optimization was also able to overcome stagnation of the traditional \HIOER{}-algorithm. In our extension, such an additional optimization is not necessary. 

We propose to replace the (non-linear and non-convex) projection operator $\Project{\gAmp}$ in reciprocal space in the \HIO{}-algorithm itself by its relaxed expression
\begin{equation}
 \OverProject{\gAmp}{\lambda_{\gAmp}} = \Identity + \lambda_{\gAmp} \left( \Project{\gAmp} - \Identity \right) \ .
 \label{eqn:DefOverrelaxationHIO}
\end{equation}
The direct space assembly in Eq. \eqref{eqn:HIORealSpaceAssembly} of the \HIO{}-algorithm remains unchanged. This corresponds to changing Eq. \eqref{eqn:HIOPolynomial} to 
\begin{equation}
 f^{(i+1)}_{\vecx} = \left[ \Identity - \Project{\ShapeRealspace} - \beta \OverProject{\gAmp}{\lambda_{\gAmp}} + \left( 1 + \beta \right) \Project{\ShapeRealspace}  \OverProject{\gAmp}{\lambda_{\gAmp}} \right] f^{(i)}_{\vecx} \equiv \HIOOOperator(\beta,\lambda_{\gAmp}) f^{(i)}_{\vecx} \ .
\label{eqn:HIOPolynomialOverrelax}
\end{equation}
If we rewrite this expression in the projectors $\Project{\ShapeRealspace}$ and $\Project{\gAmp}$, we get
\begin{equation}
 \HIOOOperator(\beta,\lambda_{\gAmp}) \equiv
 \left[ 1 + \beta \left( \lambda_{\gAmp} - 1 \right) \right] + 
 \left[ \beta - \lambda_{\gAmp} - \beta \lambda_{\gAmp} \right] \Project{\ShapeRealspace} - 
  \beta \lambda_{\gAmp}  \Project{\gAmp} +
 \left[ ( 1 + \beta )  \lambda_{\gAmp} \right]  \Project{\ShapeRealspace} \Project{\gAmp} \ .
 \label{eqn:HIOOOperator}
\end{equation}

The deviation $ \beta \left( \lambda_{\gAmp} - 1 \right) $ from the identity operator in the first term can neither be represented by the traditional \HIO{}-algorithm for any value of $\beta$ (see Eq. \eqref{eqn:HIOPolynomial}) nor by the difference map algorithm for any  ($\beta$, $\lambda_{\gAmp}$, $\lambda_{\ShapeRealspace}$) (see Eq. \eqref{eqn:DifferenceMapPolynomial}). In both cases, the previous iterative result $f^{(i)}_{\vecx}$ is weighted with $1$ or projected at least once in either direct ($\Project{\ShapeRealspace}$) or reciprocal space ($\Project{\gAmp}$) before being included in the calculation for $f^{(i+1)}_{\vecx}$.

\begin{figure}[t]
  \centering
   \subfigure[\HIOEROR{}-algorithm]{
   \includegraphics[width=0.8\textwidth]{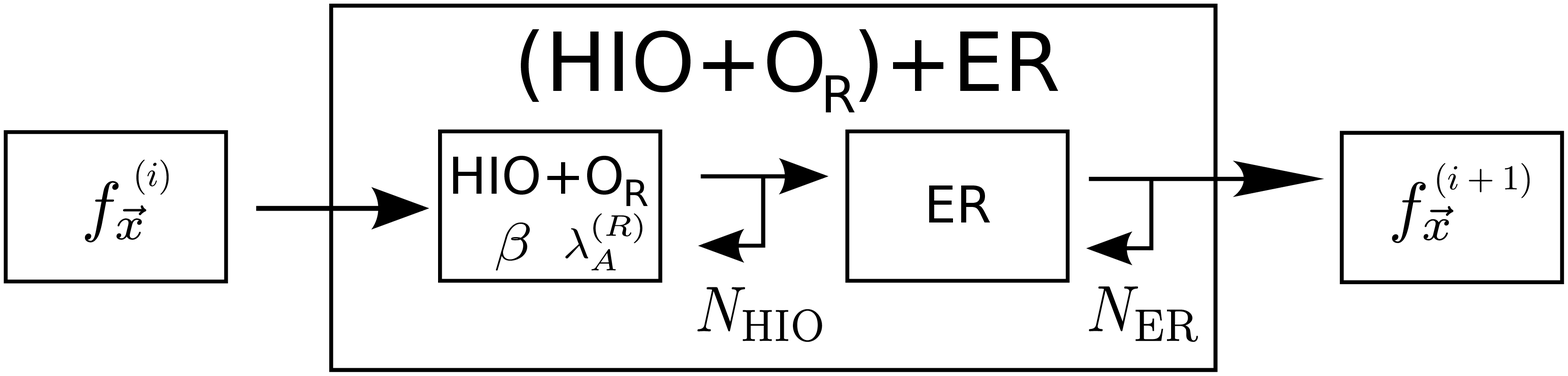} 
   \label{fig:SchematicsHIOEROComplete}
  }
  \subfigure[\ER{}-algorithm.]{
    \label{fig:SchematicsER}
    \includegraphics[height=2.4cm]{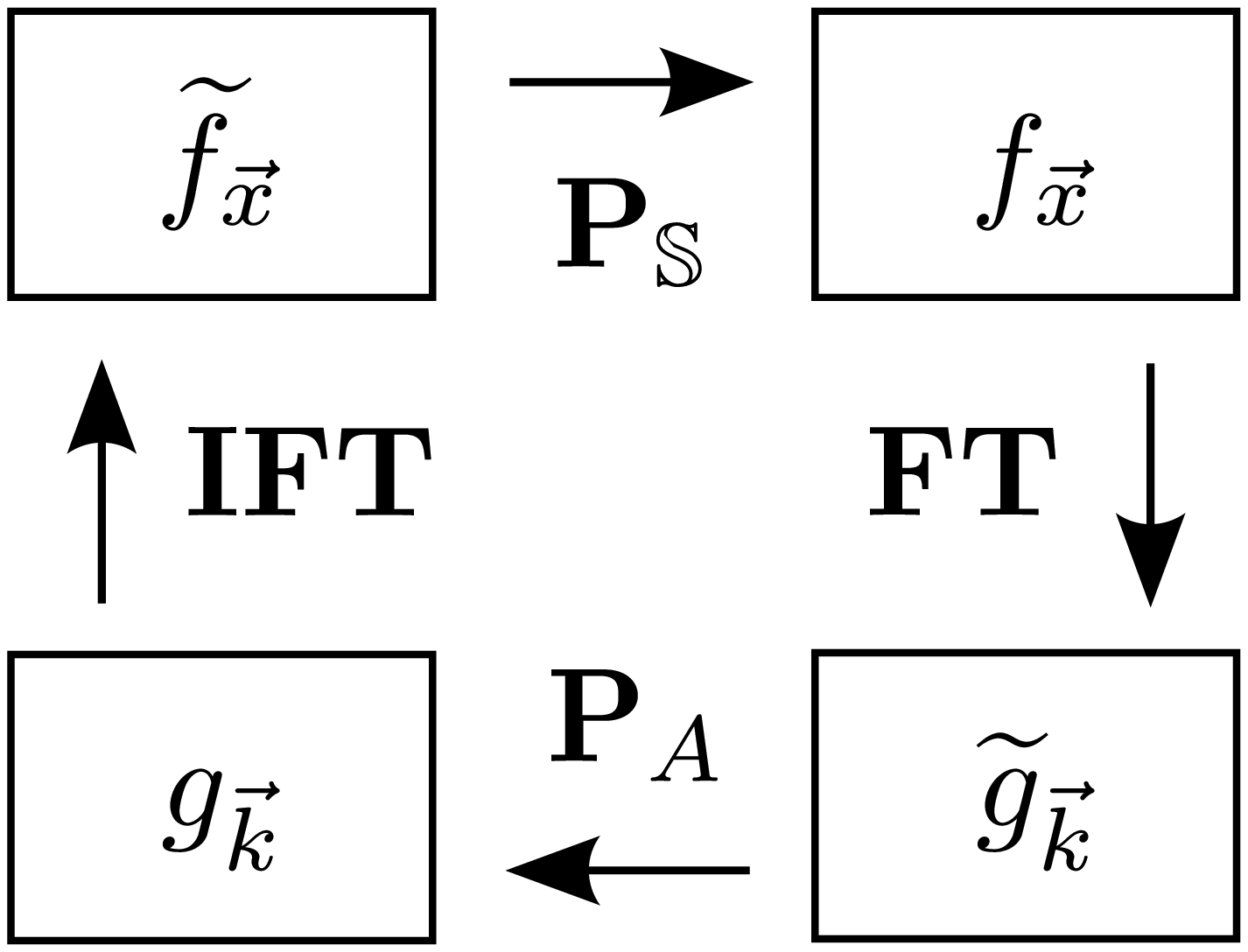} 
  } \quad
  \subfigure[\HIOOR{}-algorithm.]{
    \label{fig:SchematicsHIOO}
    \includegraphics[height=2.4cm]{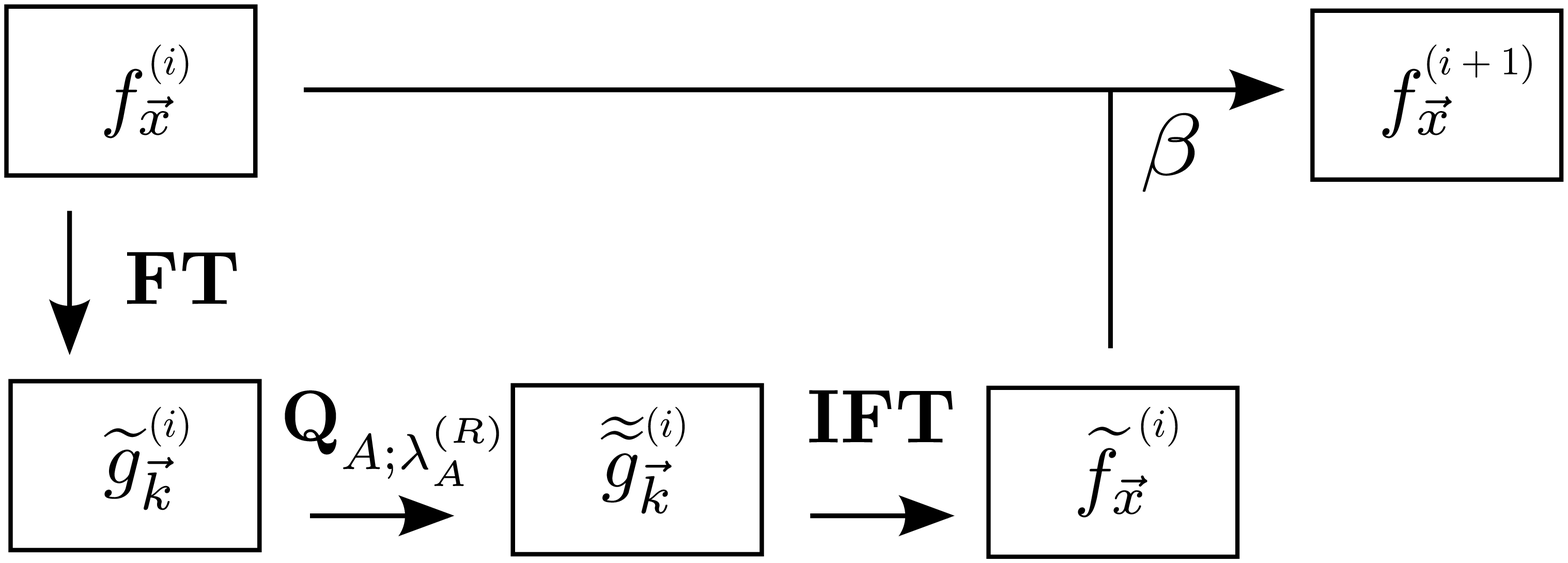} 
  }
  \caption{Graphical illustration of the \HIOEROR{}-algorithm and its building blocks.}
  \label{fig:SchematicsHIOERO}
\end{figure}

Finally, we have to choose suitable values for the additional parameter $\lambda_{\gAmp}$ in Eq. \eqref{eqn:DefOverrelaxationHIO}. If we restrict to a fixed set of values $\lambda_{\gAmp}$ for all iterations, we risk simply exchanging one trap by another or one local minimum by another. Trying to overcome local minima and stagnation, we propose a new approach: For each iteration, we \emph{randomly} select the parameter $\lambda_{\gAmp}$. More precisely, for the remainder of this paper, we investigate a uniform random distribution in the range $[1-\nu,1+\nu]$, $\nu \geq 0$, for $\lambda_{\gAmp}$ which is reassigned each iteration. Unless stated otherwise, we choose $\nu=0.5$. Note, that a fixed value $\lambda_{\gAmp} \equiv 1$ for all iterations corresponds to the usual \HIO{}-algorithm. In order to distinguish our new extension from the traditional \HIO{}-algorithm, we call our extension \HIOOR{}-algorithm. The power of this extension is strongly supported by the fact, that the \HIOOR{}-algorithm is capable of reconstructing objects without including the \ER{}-algorithm with significant success rates. If the \HIOOR{}-algorithm is combined with \ER{}, we call the algorithm \HIOEROR{}-algorithm (see Fig. \ref{fig:SchematicsHIOERO}). 

If we look at Eq. \eqref{eqn:HIOOOperator} from a bird's eye view, the most important difference of the operator $\HIOOOperator(\beta,\lambda_{\gAmp})$ to the other algorithms which have been proposed \cite{Marchesini2007Rev} is the fact, that the traps and tunnels \cite{Levi84} of the operator in general depend on $\lambda_{\gAmp}$ and, therefore, should change from iteration to iteration as a consequence of randomization. Thus, stagnation in a specific trap or tunnel is strongly reduced, as most traps and tunnels are no longer persistent throughout the iterative reconstruction process. However, the true solution $\fInput(\vecx)$ is a fixed point of the iterative procedure defined in Eq. \eqref{eqn:HIOOOperator} for \emph{all} values of $\lambda_{\gAmp}$. 

\subsection{Projection polynomials}
\label{sec:projectionPolynomials}

The extension of the \HIOER{}-algorithm to the \HIOEROR{}-algorithm is based on two ingredients: overrelaxation and randomization. In order to gain a deeper understanding of the \HIOEROR{}-algorithm, it is useful to investigate both ingredients separately. Studying the \HIO{}-algorithm extended by overrelaxation of $\Project{\gAmp}$, but without randomization is straight forward: we simply keep a fixed, predefined overrelaxation parameter $\lambda_{\gAmp}$ throughout the entire iterative reconstruction process. In this section, we present a framework for randomization of iterative projection algorithms performed in a way which is not based on overrelaxation of $\Project{\gAmp}$. Although this framework can be applied for studying randomization of any iterative projection algorithm, we limit our numerical simulations (in section \ref{sec:numericalExamples}) to parameter values whose deterministic contribution coincides the traditional \HIO{}-algorithm.

For this purpose, consider the projection polynomial operator 
\begin{subequations}
\begin{multline}
 \HProj(b,\vec{c}_{\ShapeRealspace},\vec{c}_{\gAmp}) = b \, \Identity + 
\sum_{ n = 1 }^{ n_{\text{Max}}^{\text{even}} } \left[ c_{\ShapeRealspace,2n} \left( \Project{\ShapeRealspace} \Project{\gAmp} \right)^{n} + c_{\gAmp,2n}  \left( \Project{\gAmp} \Project{\ShapeRealspace} \right)^{n}  \right] +  \\
\sum_{n = 0}^{ n_{\text{Max}}^{\text{odd}} } \left[ c_{\ShapeRealspace,2n+1} \Project{\ShapeRealspace} \left( \Project{\gAmp} \Project{\ShapeRealspace} \right)^{n} + c_{\gAmp,2n+1} \Project{\gAmp} \left( \Project{\ShapeRealspace} \Project{\gAmp} \right)^{n} \right]
 \ ,
\label{eqn:ProjectionPolynomialApproach}
\end{multline}
for obtaining the next iterative solution in the manner of Eq. \eqref{eqn:HIOPolynomial} or Eq. \eqref{eqn:HIOOOperator}. $ n_{\text{Max}}^{\text{even}}$ is given by the largest integer smaller or equal to $\frac{p}{2}$. $n_{\text{Max}}^{\text{odd}}$ is determined by the largest integer smaller or equal to $\frac{p-1}{2}$. $p$ is the maximum number of successive projection operators which should be included in $\HProj$.

This operator exploits the fact, that \emph{any} combination of projection operators $ \Project{\xi_1}^{n_1} \Project{\xi_2}^{n_2} \ldots \Project{\xi_m}^{n_m}$ of two different kinds $\xi_j \in \{ \ShapeRealspace, \gAmp \}$, $j \in \{ 1 , \ldots , m\}$, reduces to one of the four building blocks $\Project{\ShapeRealspace} \left( \Project{\gAmp} \Project{\ShapeRealspace} \right)^{n}$, $\Project{\gAmp} \left( \Project{\ShapeRealspace} \Project{\gAmp}\right)^{n}$, $\left( \Project{\ShapeRealspace} \Project{\gAmp} \right)^{n}$ and $\left( \Project{\gAmp} \Project{\ShapeRealspace} \right)^{n}$ for some integer $n\geq 0$. This is a consequence of the defining property $\Project{\xi}^{2}=\Project{\xi}$ (idempotence) of projection operators (which implies $\Project{\xi}^{n}=\Project{\xi} \ \forall \ n \geq 1$). Note, that the product of two non-commutative idempotent operators (like $\Project{\gAmp}$ and $\Project{\ShapeRealspace}$) is no longer idempotent. The special case of the identity operator $\Identity$ and a single projection operator is included for the case $n=0$ in those building blocks. However, the identity operator is included twice, namely for $n=0$ for $\left( \Project{\ShapeRealspace} \Project{\gAmp} \right)^{n}$ and $\left( \Project{\gAmp} \Project{\ShapeRealspace} \right)^{n}$. Therefore, one spurious coefficient in the projection polynomial is introduced if both building blocks are included for $n=0$. This is avoided by restricting those two building blocks to $n \geq 1$ and including the identity operator separately. Therefore, Eq. \eqref{eqn:ProjectionPolynomialApproach} represents the most general polynomial expression which contains a maximum of $q$ successive projections of two different kinds, if we incorporate the idempotence of projection operators. 

In order to guarantee, that the true solution $\fInput(\vecx)$ is a fixed point of $\HProj$ for any choice of its parameters $(b,\vec{c}_{\ShapeRealspace},\vec{c}_{\gAmp})$, it is necessary to enforce one constraint on the coefficients: If the input for the projection operators $\Project{\ShapeRealspace}$ and $\Project{\gAmp}$ coincides with the true solution $\fInput(\vecx)$, the projection operators  $\Project{\ShapeRealspace}$ and $\Project{\gAmp}$ reduce to the identity operator $\Identity$. Therefore, given the true solution $\fInput(\vecx)$ as input, the operator $ \HProj$ simplifies to the identity operator $\Identity$ for any choice of its parameters $(b,\vec{c}_{\ShapeRealspace},\vec{c}_{\gAmp})$ \emph{if and only if} the additional constraint 
\begin{equation}
 b = 1 - \sum_{n = 1}^{p} \left[ c_{n,\ShapeRealspace} + c_{n,\gAmp} \right]  \ .
\label{eqn:RestrictionPrefactorIdentityProjectionPolynomial}
\end{equation}
\label{eqn:ProjectionPolynomialApproachBothEquations}
\end{subequations}
is fulfilled. Consequently, for a maximum of $p$ successive projections in the projection polynomial $\HProj$, only $2p$ free parameters appear in $\HProj$ by incorporating the idempotence of projection operators.

Given this fix point property for the true solution $\fInput(\vecx)$, we can perform proper randomization of this operator. For this purpose, we split our coefficients $c_{\xi,n}$, $\xi \in \{ \ShapeRealspace, \gAmp \}$, $n \geq 1$, in a deterministic part $c_{\xi,n}^{(D)}$ and a random part $r_{\xi,n} \, c_{\xi,n}^{(R)}$ and assume $r_{\xi,n}$ to be uniformly distributed in the range $[-1,1]$, i.e.,
\begin{equation}
 c_{\xi,n} = c_{\xi,n}^{(D)} + r_{\xi,n} \, c_{\xi,n}^{(R)} \ .
\label{eqn:ProjPolySplittingRandomContribution}
\end{equation}
If we set some coefficients $ c_{\xi,n}^{(R)} \neq 0$ and draw statistically independent random values for $r_{\xi,n}$, we can investigate randomization of the \HIO{}-algorithm without the correlations in the coefficients $c_{\xi,n}$ introduced by overrelaxation. $b$ is fixed by Eq. \eqref{eqn:RestrictionPrefactorIdentityProjectionPolynomial} and, thus, is fully correlated to the values of $c_{\xi,n} $.

In order to obtain the traditional \HIO{}-algorithm in the limit $c_{\xi,n}^{(R)}\rightarrow 0$ for all $\xi$ and $n$, we have to choose (at least) $q=2$ (see Eq. \eqref{eqn:HIOPolynomial}). Therefore, $\HProj$ reduces to 
\begin{equation}
 \HIOROperator(b,c_{\ShapeRealspace,1},c_{\ShapeRealspace,1},c_{\gAmp,1},c_{\gAmp,2}) \equiv b \, \Identity + 
c_{\ShapeRealspace,1}   \Project{\ShapeRealspace} + 
c_{\gAmp,1}   \Project{\gAmp} +
c_{\ShapeRealspace,2} \Project{\ShapeRealspace} \Project{\gAmp} + 
c_{\gAmp,2} \Project{\gAmp} \Project{\ShapeRealspace}
\label{eqn:HIOROperator}
\end{equation}
with the additional constraint $ b = 1 - \sum_{ n=1 }^{2} \left[ c_{n,\ShapeRealspace} + c_{n,\gAmp} \right]$. The deterministic contribution $c_{\xi,n}^{(D)}$ reproduces the traditional \HIO{}-algorithm (see Eq. \eqref{eqn:HIOPolynomial}) if we set
\begin{equation}
\begin{array}{llll}
c_{\ShapeRealspace,1}^{(D)}= -1 \ , & c_{\gAmp,1}^{(D)} =-\beta \ , & 
c_{\ShapeRealspace,2}^{(D)}=1+\beta \ , & c_{\gAmp,2}^{(D)} = 0  \ .
\end{array}
\label{eqn:TraditionalHIOExpressedAsProjPoly}
\end{equation}
After replacing $\Project{\gAmp}$ by $\OverProject{\gAmp}{\lambda_{\gAmp}}$ (see Eq. \eqref{eqn:DefOverrelaxationHIO}), the  coefficients $c_{\xi,n}$ are parametrized by two parameters $\beta$ and $\lambda_{\gAmp}$ as
\begin{subequations}
\begin{equation}
\begin{array}{llll}
c_{\ShapeRealspace,1}=\beta - \lambda_{\gAmp} - \beta  \lambda_{\gAmp} \ , & c_{\gAmp,1} =-\beta \lambda_{\gAmp} \ , &
c_{\ShapeRealspace,2}=(1+\beta) \lambda_{\gAmp} \ , &  c_{\gAmp,2} = 0 \ , 
\end{array}
\end{equation}
which corresponds to
\begin{equation}
\begin{array}{llll}
c_{\ShapeRealspace,1}= - 1 - \gamma_{\gAmp} ( 1 + \beta ) \, , & c_{\gAmp,1} = -\beta ( 1+ \gamma_{\gAmp} ) \, , &
c_{\ShapeRealspace,2}=(1+\beta) ( 1 + \gamma_{\gAmp})  \, , &  c_{\gAmp,2} = 0 \, , 
\end{array}
\label{eqn:OverrelaxedHIOExpressedAsProjPoly}
\end{equation}
\end{subequations}
if we substitute $\lambda_{\gAmp} = 1 + \gamma_{\gAmp}$. $\gamma_{\gAmp}$ is uniformly distributed in $ [ - \nu , \nu ]$. The deterministic contribution in Eq. \eqref{eqn:OverrelaxedHIOExpressedAsProjPoly} is identical to Eq. \eqref{eqn:TraditionalHIOExpressedAsProjPoly} and reproduces the traditional \HIO{}-algorithm in the limit $\nu \rightarrow 0$. However, the random contribution of each coefficient $c_{\xi,n}$ is determined solely by the value of $\gamma_{\gAmp}$ and, thus, not statistically independent, but fully correlated. This is in strong contrast to a statistically independent randomization of each coefficient $c_{\xi,n}$ separately.

Consequently, we can apply the framework of projection polynomials to investigate whether the benefits of randomization rely on the precise correlations evident in Eq. \eqref{eqn:OverrelaxedHIOExpressedAsProjPoly} or if randomization of the traditional \HIO{}-algorithm implies significant benefits also in absence of these correlations.

\section{Monitoring convergence of the reconstruction}
\label{sec:Convergence}

In order to investigate the success and convergence properties of the reconstruction, we monitor three parameters. First of all, we monitor the change of the reconstructed object $\fReconstructed^{(i)}$ from iteration $(i-1)$ to iteration $(i)$. Second, we measure the mathematical distance to the (sampled) true solution $\fInput(\vecx)$. And last, but not least, we measure the deviation in reciprocal space between the a priori given reciprocal input data $\gAmp_{\veck}$ and the magnitude of the Fourier transform of the reconstructed object. All three parameters yield different information: 

The distance of the reconstructed object $ \fReconstructed^{(i)} $ after iteration $(i)$ to the true direct space object $\fInput$ is, of course, the best measure for the quality of the reconstructed image. We define this mathematical distance by an angle 
\begin{equation}
 \varphi^{(i)} = \arccos \left( \left| \braket{ \fReconstructed^{(i)} }{ \fInput } \right|  / \sqrt{\braket{ \fReconstructed^{(i)} }{ \fReconstructed^{(i)} }\braket{ \fInput }{ \fInput }} \right) \ .
 \label{eqn:angleIdealSolution}
\end{equation}
The absolute value in the nominator of the argument of the $\arccos$ eliminates the influence of the undetermined global phase in $\fReconstructed^{(i)}$. Moreover, $\varphi^{(i)}$ is identical in direct and reciprocal space (invariance of scalar product upon unitary basis transformations like the Fourier transform).

A dimensionless, normalized error measure $\epsilon^{(i)}$ which depends only on the a priori known (measured) amplitudes $\gAmp_{\veck}$ and not on the true direct space solution $\fInput(\vecx)$ is
\begin{equation}
\epsilon^{(i)}  = \frac{ \braket{ |\gReconstructed^{(i)}| - \gAmp } { |\gReconstructed^{(i)}| - \gAmp } } { \braket{ \gAmp }{ \gAmp } }  = \frac{1}{\left\| \gAmp \right\|_{2}^{2}} \ \sum_{\veck} \left( |\gReconstructed^{(i)}(\veck)| - \gAmp(\veck) \right)^2 \ ,
 \label{eqn:DefErrorMetrics}
\end{equation}
where $ \left\| \cdot \right\|_p$ is the $p$-norm. Since the construction of the absolute value is a non-linear, non-invertible map, the result can in general be quite different from the angle defined above. In fact, two quite different direct space objects $\fInput$ (hence quite different complex Fourier transforms) may posses an extremely similar distribution of the magnitude $\gAmp$ in reciprocal space \cite{Seldin1990,PhysRevB.78.174110}.

In addition, from a numerical point of view, the change of the reconstructed object $\fReconstructed^{(i)}$ from iteration to iteration is very important, too. This must not be confused with the change of the non-invertible map $\epsilon^{(i)} $ from iteration to iteration: Moving along a connected line of constant error metric does not change the error metric itself, but the reconstructed object $\fReconstructed^{(i)}$ may change arbitrarily. Nevertheless, the algorithm may converge (i.e. no change from iteration to iteration), but not to the true solution $\fInput$ of the problem \cite{PhysRevB.78.174110,Levi84}. For simulated data, we can judge if the algorithm converged to the true solution $\fInput$ by Eq. \eqref{eqn:angleIdealSolution}. A suitable choice for observing the change from iteration to iteration is given by the angle
\begin{equation}
 \chi^{(i)} = \arccos \left( \left| \braket{ \fReconstructed^{(i-1)} }{ \fReconstructed^{(i)} } \right|  / \sqrt{\braket{ \fReconstructed^{(i-1)} }{ \fReconstructed^{(i-1)} }\braket{ \fReconstructed^{(i)} }{ \fReconstructed^{(i)} }} \right) \ .
 \label{eqn:angleConvergence}
\end{equation}
In contrast, the $p$-norm $ \delta^{(i)} = \left\|\fReconstructed^{(i-1)}-  \fReconstructed^{(i)}\right\|_p $ of the difference is not optimal for measuring the convergence of the reconstruction process, because it does not eliminate global phase shifts from iteration $(i-1)$ to iteration $(i)$.

\section{Numerical examples}
\label{sec:numericalExamples}

\begin{figure}[t]
  \centering
  \subfigure[This object belongs to Eq. \eqref{eqn:DefPurePhaseObject1}.]{
    \label{fig:OrigPurePhaseObject1}
    \includegraphics[width=0.35\textwidth]{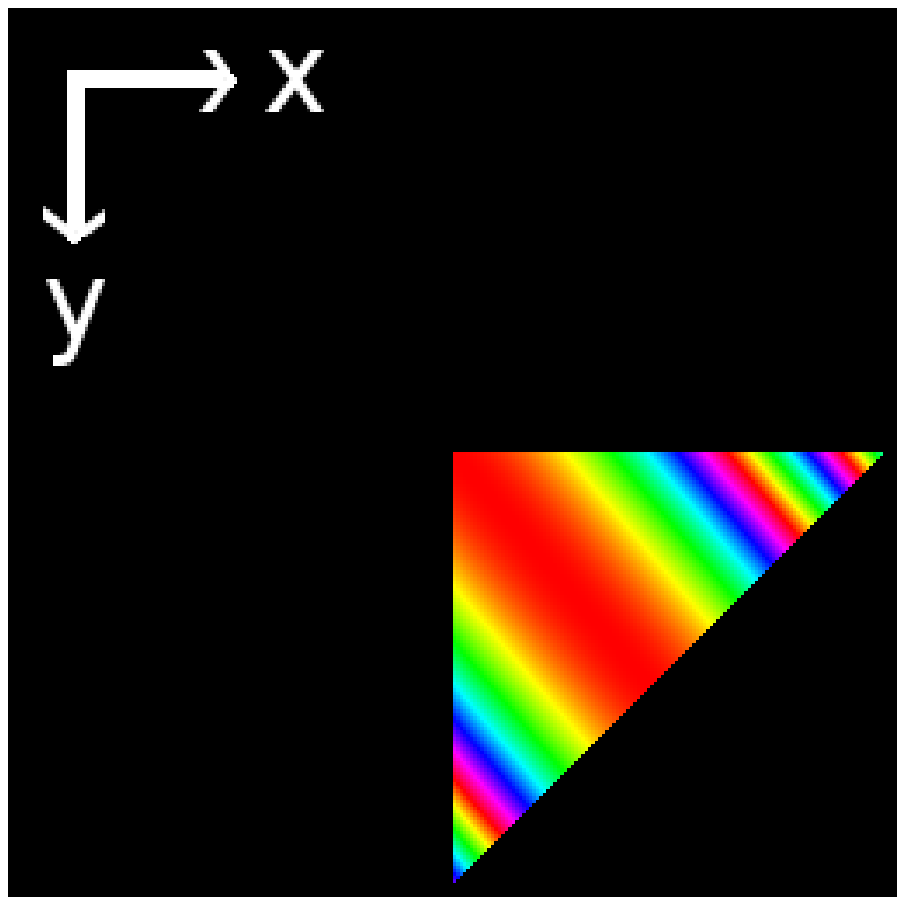} 
  } \quad
  \subfigure[This object belongs to Eq. \eqref{eqn:DefPurePhaseObject2}.]{
    \label{fig:OrigPurePhaseObject2}
    \includegraphics[width=0.35\textwidth]{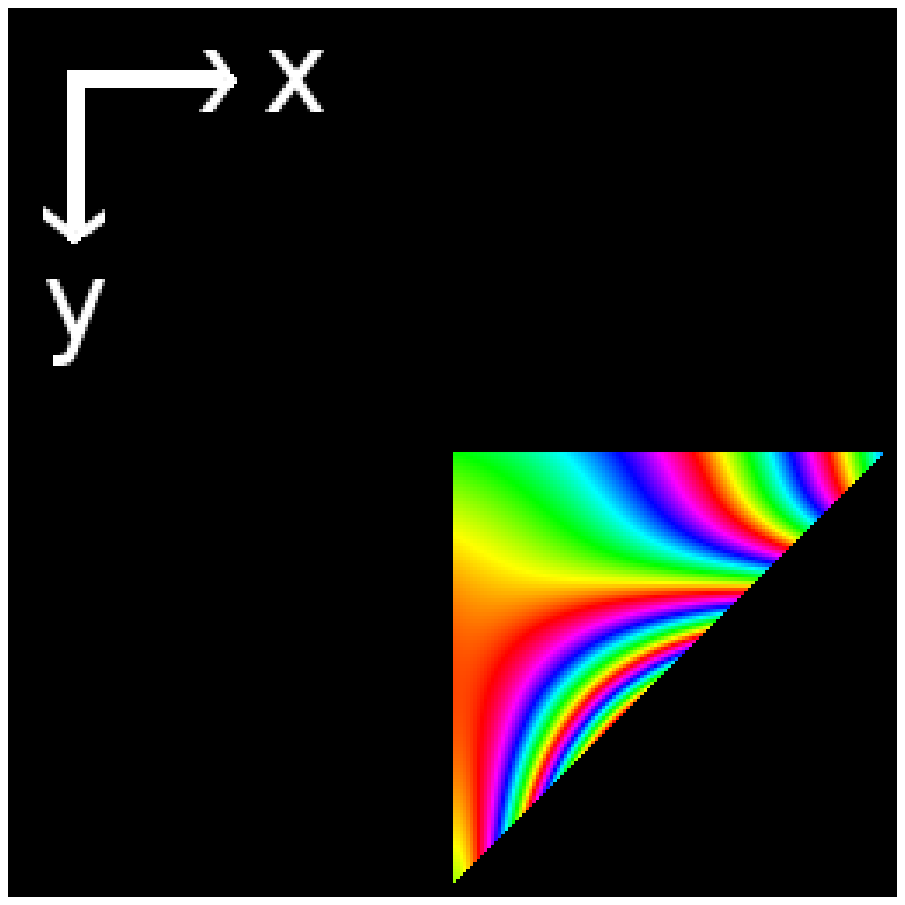} 
  }
  \caption{Pure phase objects used for the investigation of the convergence of the \HIOEROR{}-algorithm. The magnitude of both objects is constant. The phase is plotted using a HSV color-bar, however, the region outside the shape $\ShapeRealspace$ has been set to black. The oversampling ratio is $\sigma=8.456$.}
  \label{fig:PurePhaseObjectsRealSpace}
\end{figure}

\begin{figure}[t]
 \centering
 \subfigure[Magnitude of the purely real object which is used as $\fInput$ for the investigation of the convergence of the \HIOEROR{}-algorithm.]{
  \includegraphics[width=0.41\textwidth]{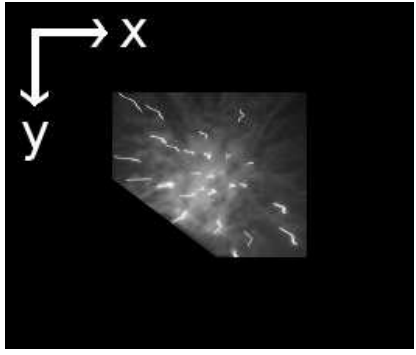}
   \label{fig:PureRealObjectRealSpaceFireworksInput}
 } \quad
 \subfigure[Results for the success rate of the \HIOER{} and the \HIOEROR{}-algorithm for the object depicted in Fig. \subref{fig:PureRealObjectRealSpaceFireworksInput}. The parameter $\beta$ was fixed to $0.8$.]{
   \includegraphics[width=0.49\textwidth]{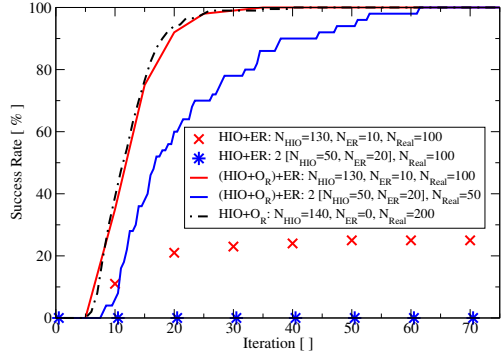}
   \label{fig:PureRealObjectRealSpaceFireworksSuccessRateResult}
 }
 \caption{Investigation of the \HIOEROR{}-algorithm for a purely real object $\fInput$ with strong variation of its magnitude over short length scales (Oversampling ratio $\sigma=5.06$). In Fig. \subref{fig:PureRealObjectRealSpaceFireworksSuccessRateResult}, continuous lines represent the \HIOEROR{}-algorithm, isolated dots the \HIOER{}-algorithm. A pure \HIOOR{}-calculation without \ER{} is included as black, dash-dotted curve.}
 \label{fig:PureRealObjectRealSpaceFireworks}
\end{figure}

In this section, we demonstrate the improvements resulting from randomized overrelaxation for three carefully chosen numerical examples: First, we consider two pure phase objects 
\begin{equation}
 f(\vecx) = \exp\left( \imagUnit \xi(\vecx) \right) \ShapeFunction_{\ShapeRealspace}(\vecx) \ ,
\label{eqn:DefPurePhaseObject}
\end{equation}
where $\xi(\vecx)$ is a \emph{real} function and $\ShapeFunction_{\ShapeRealspace}(\vecx)$ is the shape function
\begin{equation}
 \ShapeFunction_{\ShapeRealspace}(\vecx) = \left\{ 
\begin{array}{ll}
 0 & \text{ if } \vecx \notin \ShapeRealspace \ , \\
 1 & \text{ if } \vecx \in \ShapeRealspace \ .
\end{array}
\right. 
\end{equation}
The reconstruction of the first phase object (see Fig. \ref{fig:OrigPurePhaseObject1}) is plagued by (at least) one strong local minimum far away from the true solution $\fInput$ which results in severe problems for the traditional \HIOER{}-algorithm. The reconstruction of the second phase object (see Fig. \ref{fig:OrigPurePhaseObject2}) is plagued by many local minima near the perfect solution $\fInput$, in particular, by stripes \cite{Fienup1986b}. Both phase objects have a smooth variation of their phase, i.e., phase variations of $2 \pi$ extend over several sampling points. As a third test object, we investigate a purely real object $\fInput(\vecx)$ with strong variation of its magnitude over very short length scales and a different shape (see Fig. \ref{fig:PureRealObjectRealSpaceFireworksInput}). In fact, the success rate of the traditional \HIO{}-algorithm without randomized overrelaxation was lowest for this third example. From the numerical results, we conclude that randomized overrelaxation is a powerful extension for a wide class of objects and not only for objects possessing some particular features.

The first pure phase objects which we used in our numerical simulations is given by 
\begin{equation}
 \xi_1(x,y) = \left( 2 \pi \right)^2 \left[ \left( \frac{x}{b_1} \right)^2 + \left( \frac{y}{c_1} \right)^2 \right] \ ,
 \label{eqn:DefPurePhaseObject1}
\end{equation}
where $\xi_1$ is defined according to Eq. \eqref{eqn:DefPurePhaseObject} and the parameters were chosen as $b_1=1.5515$ and $c_1=-1.835$. The shape $\ShapeRealspace$ has been restricted to the triangular region defined by $(0,0)$, $(0,a)$ and $(a,0)$ with $a=0.97$. The object has been sampled on an equidistant grid with $N_1 \times N_2=256 \times 256$ data points in the interval $(x,y) =(-1,-1)$ to $(x,y)= (1,1)$. The resulting object is depicted in Fig. \ref{fig:OrigPurePhaseObject1}.

The second object is defined by 
\begin{equation}
  \xi_2(x,y) = \left( 2 \pi \right) \left[ 
  \left( x - b_2 \right)^3 + \left( y - c_2 \right)^2 + \frac{x^2 y^3 }{c_2^2 b_2^3} 
 \right] 
 \label{eqn:DefPurePhaseObject2}
\end{equation}
with the same shape $\ShapeRealspace$ and sampling parameters as in the first example. The parameters were set to $b_2=-0.3515$ and $c_2=0.535$. The phase field in this case is depicted in Fig. \ref{fig:OrigPurePhaseObject2}.

A reconstruction is considered successful once the angle $\varphi^{(i)}$ to the input object (see Eq. \eqref{eqn:angleIdealSolution}) falls below some given limit $\varphi_{\mathrm{Converged}}$ (=$0.05^{\circ}$ unless stated otherwise). We repeat the reconstruction process for $\NReal$ initial trials (random phases at each $\veck$-point, amplitudes $\gAmp^{(0)}_{\veck}$ equal to given amplitudes $\gAmp_{\veck}$). Finally, we calculate the success rate which tells us the percentage of reconstruction processes that have been successful up to iteration $(i)$. A good reconstruction algorithm 
\begin{enumerate}[(i)]
 \item should reach a success rate of almost $100\%$,
 \item should not depend on the starting point (as long as no good starting point is available),
 \item should perform the reconstructions with little computational effort and
 \item should possess these properties for a wide range of its internal parameters (i.e., $\beta$, $\nu$, $N_{\mathrm{\HIO}}$ and $N_{\mathrm{\ER}}$ in case of the \HIOEROR{}-algorithm).
\end{enumerate}

\begin{figure}[t]
  \centering
  \subfigure[First phase object defined in Eq. \eqref{eqn:DefPurePhaseObject1}.]{
    \label{fig:OrigPurePhaseObject1Comparison}
    \includegraphics[width=0.48\textwidth]{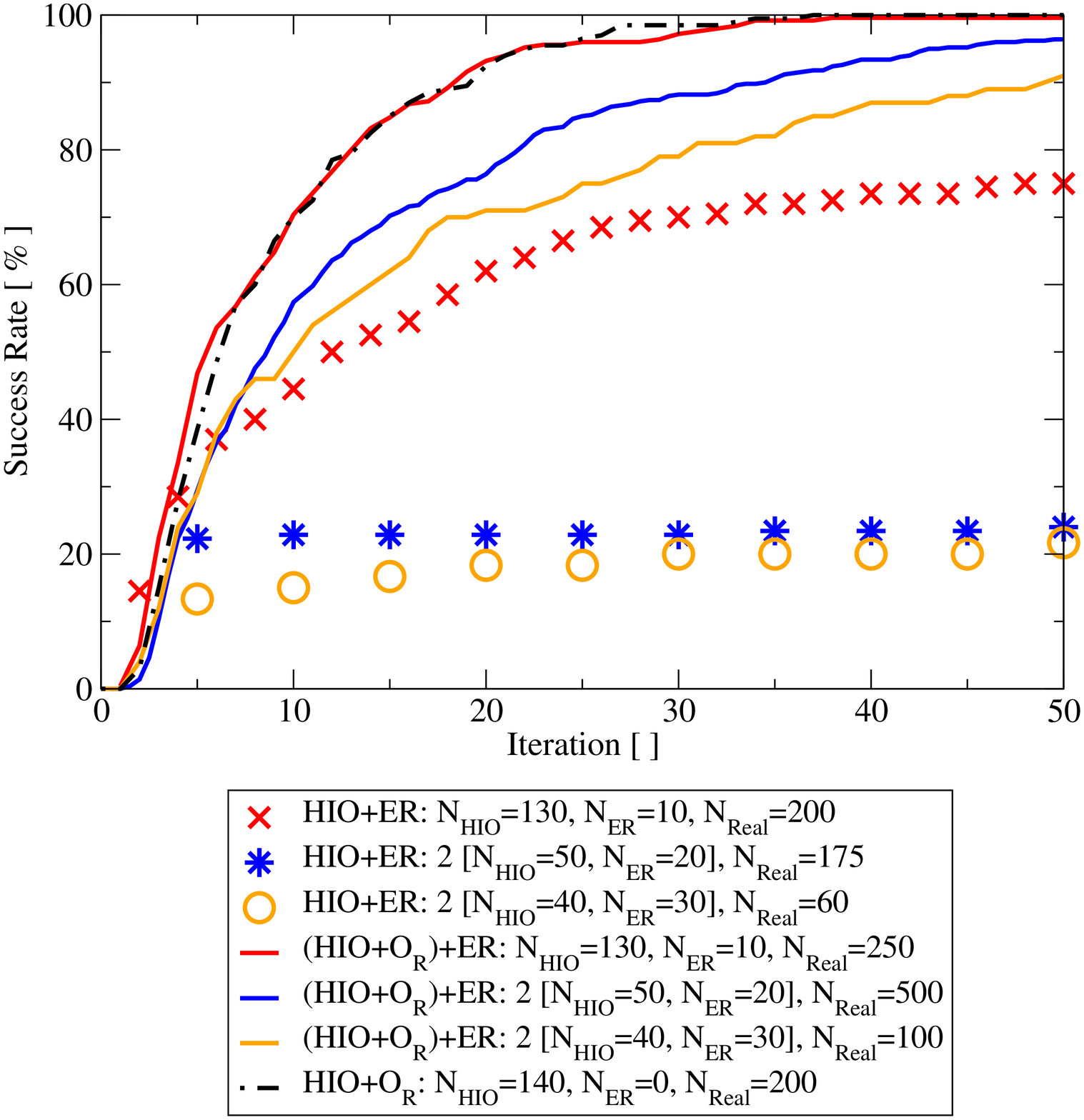} 
  }  \ 
  \subfigure[Second phase object defined in Eq. \eqref{eqn:DefPurePhaseObject2}.]{
    \label{fig:OrigPurePhaseObject2Comparison}
    \includegraphics[width=0.48\textwidth]{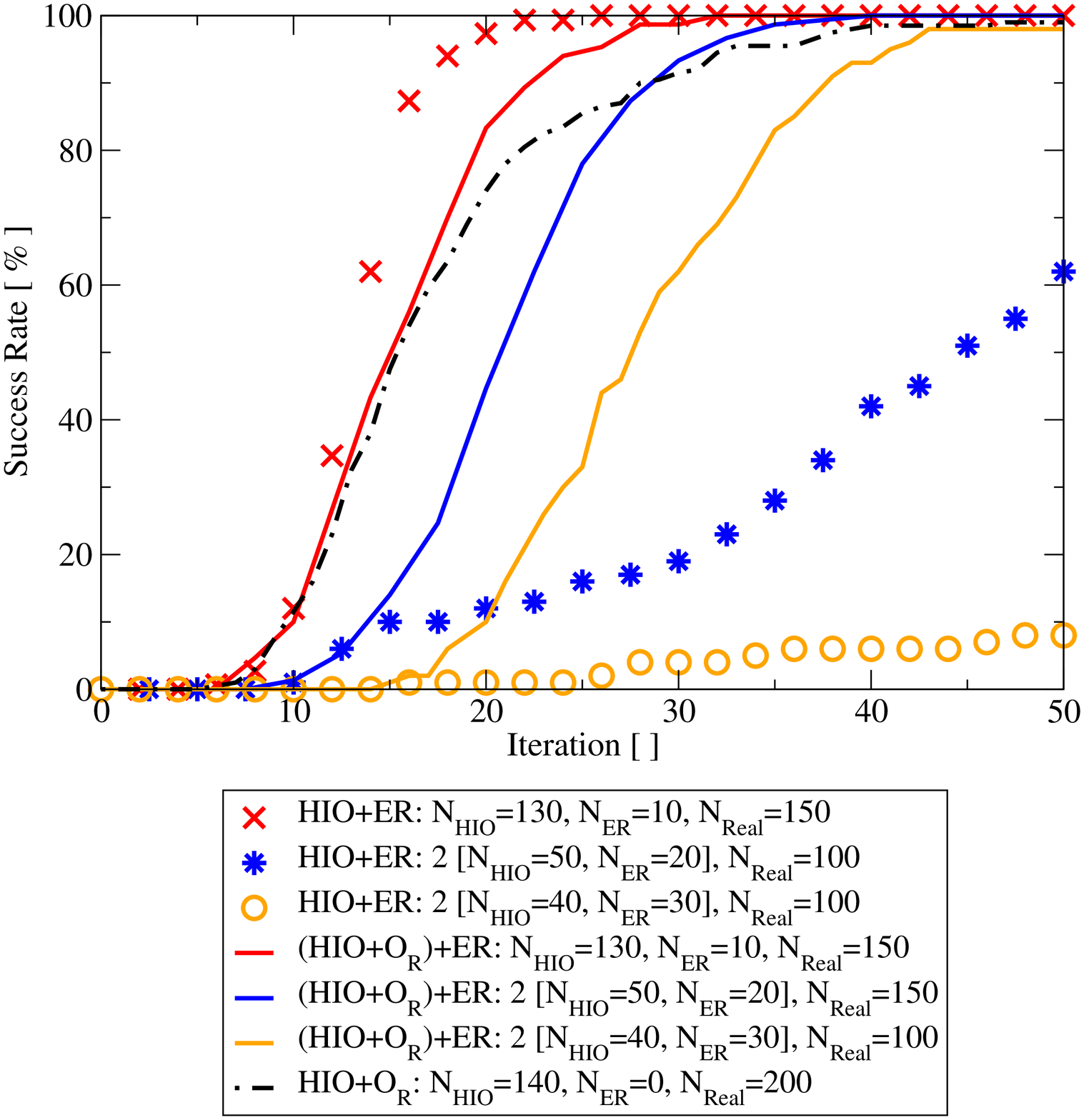} 
  }
  \caption{Comparison of the success rate of reconstructions of pure phase objects (see Eq. \eqref{eqn:DefPurePhaseObject}, \eqref{eqn:DefPurePhaseObject1} and \eqref{eqn:DefPurePhaseObject2}) with the \HIOER{}- and the \HIOEROR{}-algorithm. The parameter $\beta$ was fixed to $0.85$. Continuous lines represent results of the \HIOEROR{}-algorithm, isolated dots of the \HIOER{}-algorithm. A pure \HIOOR{}-calculation without \ER{} is included as black, dash-dotted curve.}
  \label{fig:PurePhaseObjectsSuccessRate}
\end{figure}

\begin{figure}[t]
  \centering
  \includegraphics[width=0.85\textwidth]{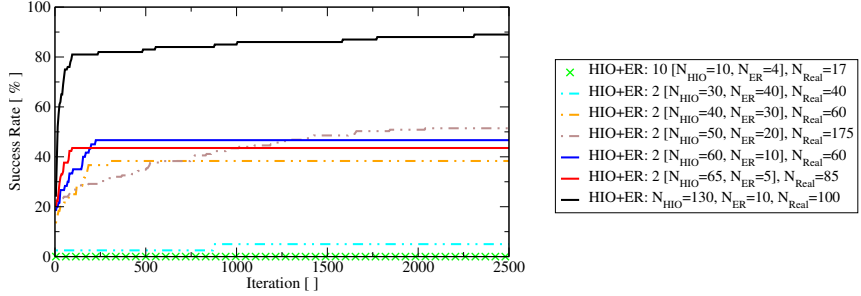}  
  \caption{Long-term stagnation of the success rate of the traditional \HIO{}-algorithm (without overrelaxation and without randomization) for the first phase object (see Eq. \ref{eqn:DefPurePhaseObject1}) for different choices of the internal parameters $(N_{\text{HIO}},N_{\text{ER}})$, but fixed $\beta=0.85$.}
  \label{fig:OrigPurePhaseObject1StagnationHIOER}    
\end{figure}

\begin{figure}[t]
 \centering
 \subfigure[Dependence of the success rate on the bounds of the uniformly distributed relaxation parameter $\lambda_{\gAmp}$.]{
  \includegraphics[width=0.48\textwidth]{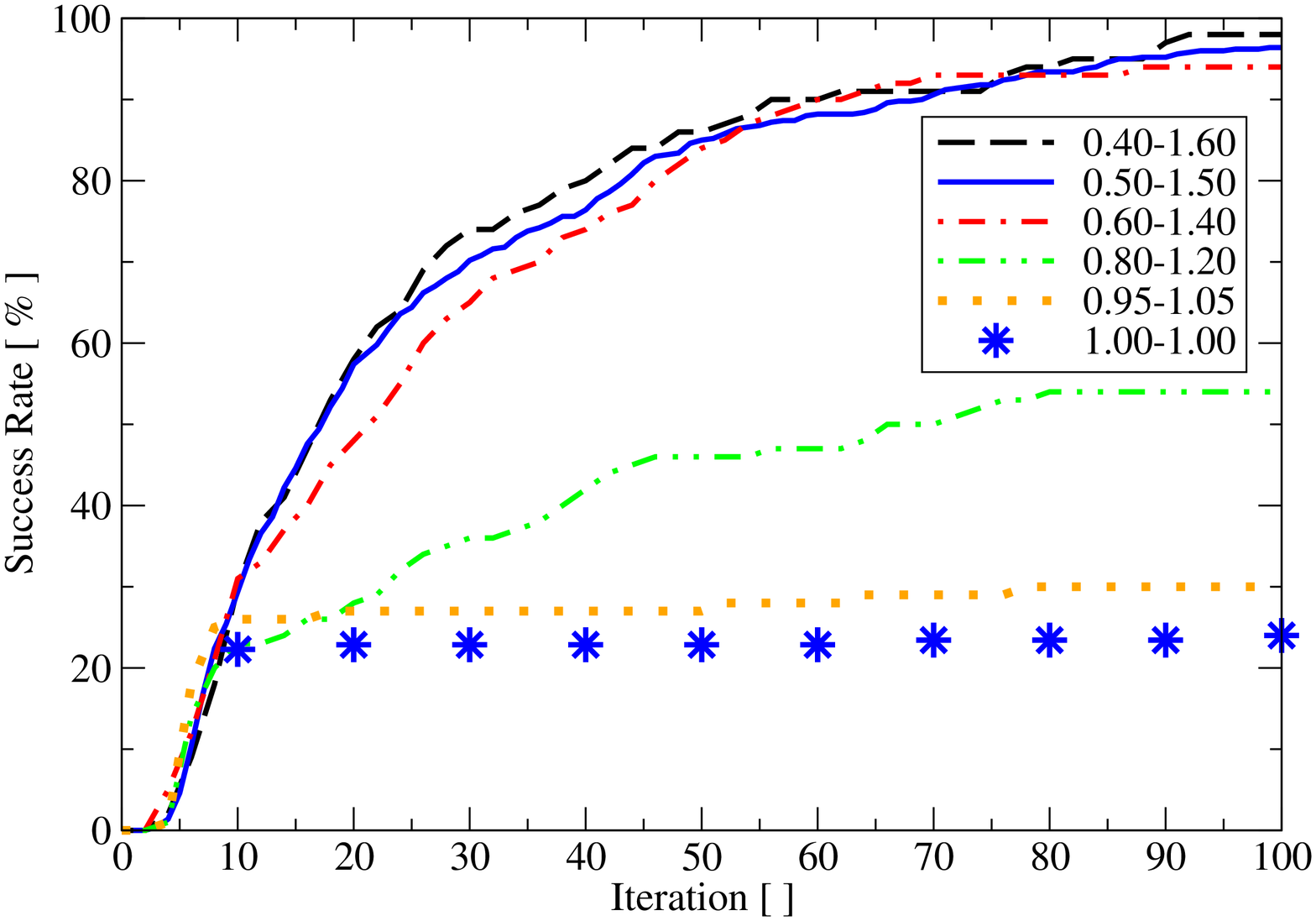}
   \label{fig:randomizationLimitsHIOEROR}
 } \ 
 \subfigure[Dependence of the success rate on the parameter $\beta$ of the \HIOEROR{}-algorithm.]{
  \includegraphics[width=0.48\textwidth]{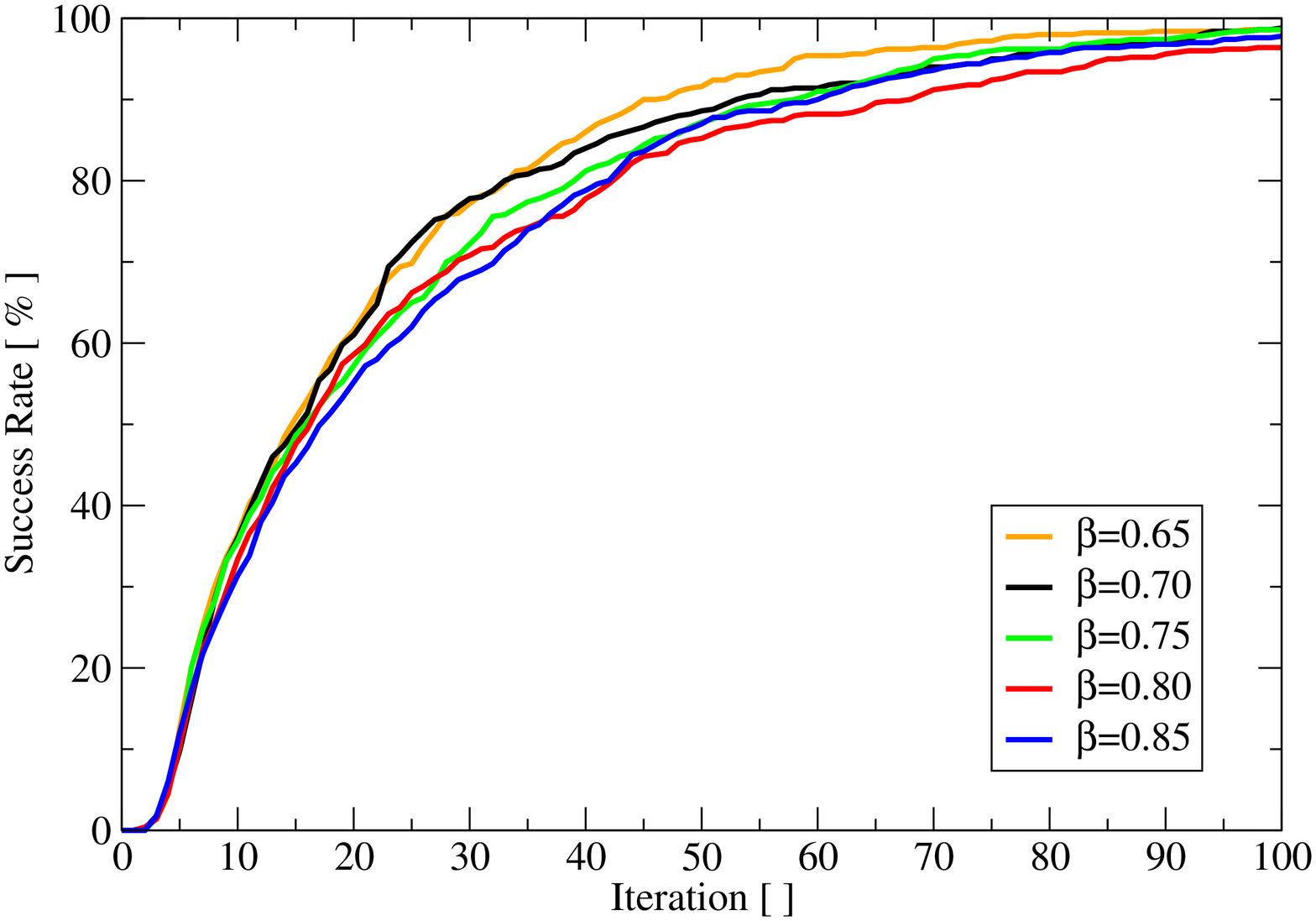} 
 \label{fig:OrigPurePhaseObject1DependenceBeta}
 }
 \caption{Investigation of the sensitivity of the \HIOEROR{}-algorithm with $N_{\text{HIO}}=50$ and $N_{\text{ER}}=20$ on the choice of the parameter $\beta$ and on the range of the uniform distribution determining the relaxation parameter $\lambda_{\gAmp}$ for the first phase object (see Eq. \ref{eqn:DefPurePhaseObject1}).}
 \label{fig:ComparisonProjectionPolynomialsWithHIOO}
\end{figure}

\begin{figure}[t]
 \centering
 \subfigure[First phase object defined in Eq. \eqref{eqn:DefPurePhaseObject1}.]{
   \includegraphics[width=0.46\textwidth]{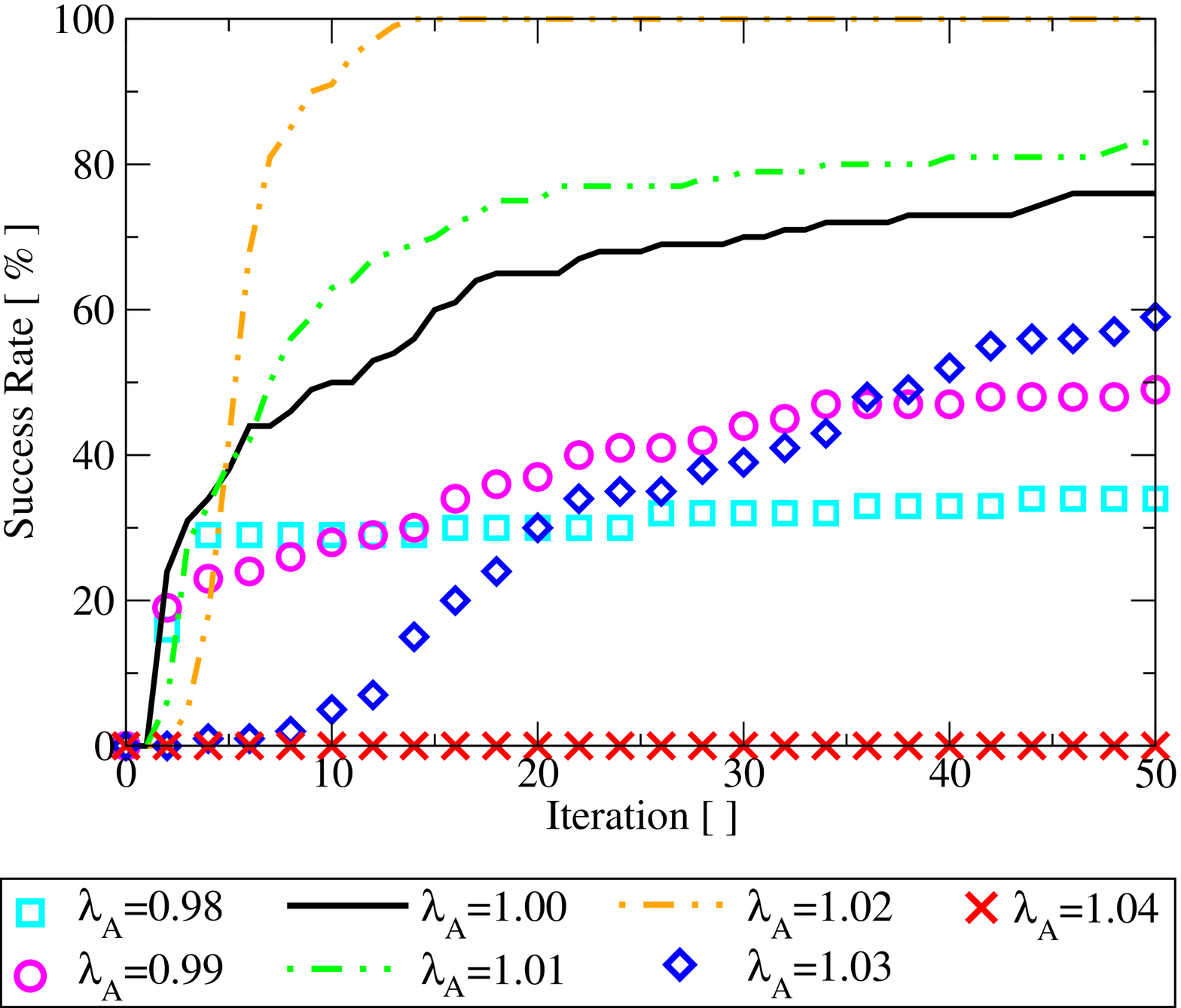}
 } \ 
 \subfigure[Second phase object defined in Eq. \eqref{eqn:DefPurePhaseObject2}.]{
   \includegraphics[width=0.46\textwidth]{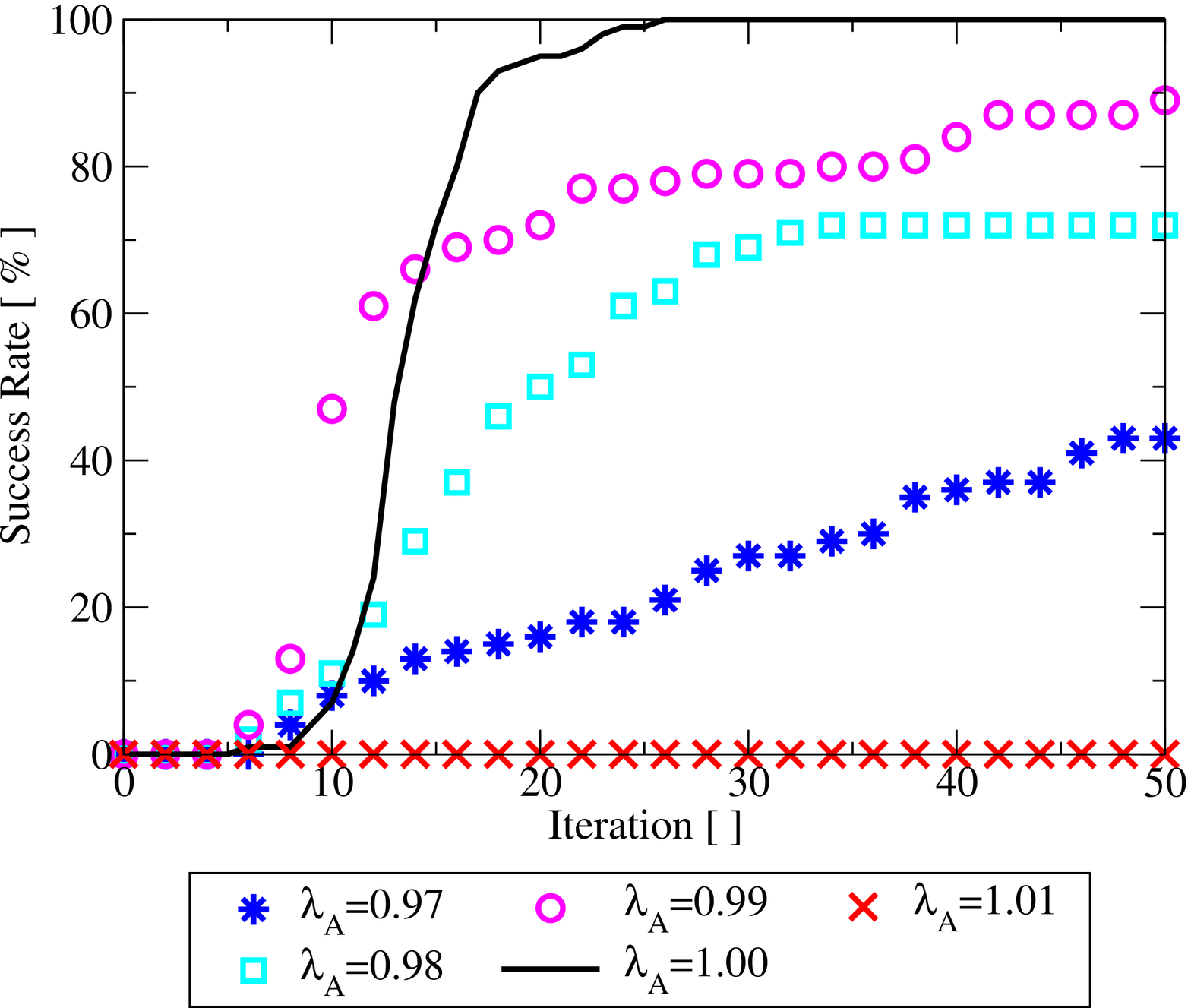}
 }
 \caption{Success rate of the \HIOER{}-algorithm for fixed overrelaxation $\lambda_{\gAmp}$ (no randomization). Parameters are chosen as ($N_{\text{HIO}}=130$,$N_{\text{ER}}=10$) and $\beta=0.85$.}
 \label{fig:ComparisonNoRandomizationPureOverrelaxation}
\end{figure}
   
Each iteration in the \HIOER{}-algorithm and its \HIOEROR{}-extension can be implemented very efficiently exploiting the good scaling properties of the FFT-algorithm. Therefore, as long as the number of iterations is not too high, the third requirement is clearly met by both, the \HIOER{} and \HIOEROR{}-algorithm.

In order to investigate the other points, we plotted the respective success rates for the \HIOER{}- and \HIOEROR{}-algorithm for different internal parameters in Fig. \ref{fig:PurePhaseObjectsSuccessRate}. We normalized the computational effort, i.e., one iteration of the \HIOEROR{}-algorithm with $N_{\mathrm{HIO}}=130$ and $N_{\mathrm{ER}}=10$ corresponds to two successive iterations with $N_{\mathrm{HIO}}=50$ and $N_{\mathrm{ER}}=20$. This prefactor of $2$ can be found in the legend of the graph in front of the square brackets for each calculation. 

The inclusion of overrelaxation improves the success rate for the first phase object defined by Eq. \eqref{eqn:DefPurePhaseObject1} for each combination of $N_{\mathrm{HIO}}$ and $N_{\mathrm{ER}}$: in contrast to the \HIO{} based calculations, it quickly converged to $100\%$. In fact, overrelaxation is so powerful that stagnation in the \HIO{}-algorithm is eliminated entirely and intermediate \ER{}-iterations are no longer necessary (see black dash-dotted curve in Fig. \ref{fig:PurePhaseObjectsSuccessRate}). The success rate of the traditional \HIOER{}-algorithm stagnates on some level far below $100\%$. This stagnation is significant in the long term behavior, as we illustrate in Fig. \ref{fig:OrigPurePhaseObject1StagnationHIOER} by plotting the success rate of the traditional \HIOER{}-algorithm for 50 times as many iterations as in Fig. \ref{fig:OrigPurePhaseObject1Comparison}. Note, that all trials that did not converge after 2500 iterative steps using the traditional \HIOER{}-algorithm with $N_{\mathrm{HIO}}=130$ and $N_{\mathrm{ER}}=10$ in Fig. \ref{fig:OrigPurePhaseObject1StagnationHIOER} are separated from the true solution by an angle greater than $\varphi=55^{\circ}$.

If we consider the second phase object defined by Eq. \eqref{eqn:DefPurePhaseObject2}, the traditional \HIOER{}-algorithm performs better here than for the first phase object. For the one specific set of parameters $N_{\mathrm{HIO}}=130$ and $N_{\mathrm{ER}}=10$, the success rate of the traditional \HIOER{}-algorithm is even slightly superior to the \HIOEROR{}-extension, but for this specific set of parameters, the success rate converges to $100\%$ very quickly for both algorithms. However, for the set of parameters ($N_{\mathrm{HIO}}=50$, $N_{\mathrm{ER}}=20$) and ($N_{\mathrm{HIO}}=40$, $N_{\mathrm{ER}}=30$), the success rate quickly converges to $100\%$ only for the \HIOEROR{}-algorithm, but not for the \HIOER{}-algorithm. Hence, the \HIOEROR{}-algorithm is much more stable with respect to the particular choice of the internal parameters $N_{\mathrm{HIO}}$ and $N_{\mathrm{ER}}$ including, in particular, the case $N_{\mathrm{ER}}=0$. 

The traditional \HIOER{}-algorithm also has severe problems performing a successful reconstruction for the real object depicted in Fig. \ref{fig:PureRealObjectRealSpaceFireworksInput}: The success rate for the reconstruction process is very poor if we apply our criterion $\varphi_{\mathrm{Converged}} \leq 0.05^{\circ}$ to determine successful reconstructions. The \HIOEROR{}-algorithm is much more successful and reaches success rates of $100\%$ with few iterations (see Fig. \ref{fig:PureRealObjectRealSpaceFireworksSuccessRateResult}). Furthermore, reaching the success rate of $100\%$ does not sensitively depend on the internal parameters $N_{\mathrm{HIO}}$ and $N_{\mathrm{ER}}$. Again, keeping $N_{\mathrm{ER}}$ small compared to $N_{\mathrm{HIO}}$ or eliminating \ER{}-iterations completely yields the best results. Keep in mind, that neither the reality constraint nor the positivity constraint of the object have been exploited during reconstruction.

For the second and third test object, all trials that failed to converge to the true solution $\fInput$ basically evolved towards the correct direction. However, at some point, stagnation in form of stripes close to the true solution $\fInput$ appeared. These stripes are very persistent and prevent the usual \HIOER{}-algorithm from further convergence to the true solution $\fInput$. Typically, the mathematical distance $\varphi$ of the objects containing stripes to the true solution $\fInput$ is in the order of $0.05$ to $5.0$ degrees. In case of the purely real test object, after 150 iterations all reconstructed objects reached a distance $\varphi \leq 0.5^{\circ} $ ($= 10 \cdot \varphi_{\mathrm{Converged}}$) for $\text{\HIO}_{130}-\text{\ER}_{10}$ and 31\% for $\text{\HIO}_{50}-\text{\ER}_{20}$. The remaining 71\% of the objects for $\text{\HIO}_{50}-\text{\ER}_{20}$ reached a distance $\varphi \leq 5^{\circ}$ ($= 100 \cdot \varphi_{\mathrm{Converged}}$) using traditional \HIOER{}-algorithm and 150 iterations.  Note, that the change $\chi^{(i)}$ (see Eq. \eqref{eqn:angleConvergence}) from iteration to iteration does not vanish, i.e. numerically, the algorithm did not converge within the given number of iterations. 

Figure \ref{fig:randomizationLimitsHIOEROR} shows the behavior of the success rate of the \HIOEROR{}-algorithm for different values of the parameter $\nu$. In addition, Fig. \ref{fig:OrigPurePhaseObject1DependenceBeta} depicts the success rate for different values of the parameter $\beta$. Within the range $\beta \in [0.5, 1.0]$, which is typically used in the traditional \HIOER{}-algorithm, the \HIOEROR{}-extension does not sensitively depend on the particular choice of $\beta$. Therefore, the \HIOEROR{}-algorithm does not depend sensitively on any of its internal parameters $N_{\text{HIO}}$, $N_{\text{ER}}$, $\beta$ and $\nu$. Hence, it fulfills all four requirements on a good reconstruction algorithm which we stated above. 

Two concepts (overrelaxation and randomization) have been exploited to modify the \HIO{}-algorithm: Fig. \ref{fig:ComparisonNoRandomizationPureOverrelaxation} depicts the success rate, if overrelaxation is performed with a static parameter $\lambda_{\gAmp}$. We see extremely sensitive behavior of the success rate on the precise value of $\lambda_{\gAmp}$ if it is kept fixed. Whereas increasing $\lambda_{\gAmp}$ up to $1.02$ improved the success rate for the first object, it completely vanished for $\lambda_{\gAmp}=1.01$ for the second phase object. For a deviation from $1.00$ greater than ten percent, almost no successful reconstructions have been observed for any of the three test objects. In addition, we do not a priori know the (non-universal) optimum value for $\lambda_{\gAmp}$ which depends on the object $\fInput$. Consequently, static overrelaxation does not fulfill the requirements for a good reconstruction algorithm stated above.

\begin{figure}
\centering
   \includegraphics[width=0.99\textwidth]{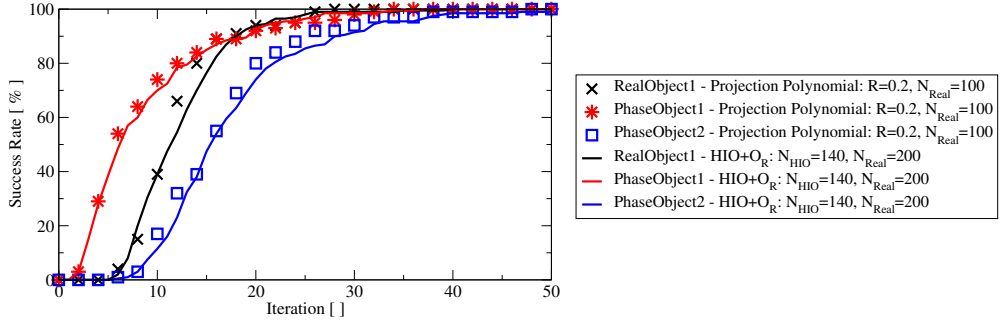}
   \caption{Comparison of the success rate of both frameworks which provide a generalization of the traditional \HIO{}-algorithm based on randomization. Continuous lines illustrate the behavior for randomized overrelaxation of $\Project{\gAmp}$ (see Eq. \eqref{eqn:HIOPolynomialOverrelax}), whereas dots represent the behavior of the success rate resulting from independent randomization of the coefficients in a projection polynomial (see Eq. \eqref{eqn:ProjPolySplittingRandomContribution} and Eq. \eqref{eqn:HIOROperator}). In both cases, the deterministic contribution is equivalent to the traditional \HIO{}-algorithm with parameters $N_{\text{HIO}}=140$ and $\beta=0.85$. No \ER{} has been performed.}
   \label{fig:PureRandomizationWithoutOverrelaxation}
\end{figure}

Based on the discussion in sec. \ref{sec:projectionPolynomials}, we test randomization of the \HIO{}-algorithm for $\beta=0.85$ without employing overrelaxation by setting the amplitudes $c_{\xi,n}^{(R)}$ for the random contribution to the coefficients $c_{\xi,n}$ in the operator $\HIOROperator$ defined in Eq. \eqref{eqn:HIOROperator} equal to $ c_{\ShapeRealspace,1}^{(R)}= c_{\gAmp,1}^{(R)} = c_{\ShapeRealspace,2}^{(R)} = c_{\gAmp,2}^{(R)}= R = 0.2 $. The deterministic contribution $c_{\xi,n}^{(D)}$ is set according to Eq. \eqref{eqn:TraditionalHIOExpressedAsProjPoly}. The result for all three test objects is illustrated in Fig. \ref{fig:PureRandomizationWithoutOverrelaxation} and compared to the result for the \HIOOR{}-algorithm. Clearly, pure randomization without overrelaxation performs equally well as the \HIOOR{}-algorithm for all three test objects. However, in our opinion, the \HIOOR{}-algorithm should be preferred over randomization without overrelaxation, because (i) it requires only two Fourier transformations for each iteration (lower computational effort) and (ii) it has less degrees of freedom (one uniform random distributions instead of four). Note, however, that due to the large number of degrees of freedom of projection polynomials, further optimization of this approach is possible (including more advanced statistical correlations in the coefficients of the projection polynomial). In addition, more elaborate random distributions might improve both approaches further. 

In summary, we succeeded in overcoming stagnation of the traditional \HIOER{}-algorithm for various objects. For this purpose, the concept of overrelaxation was applied to the reciprocal space projection in the \HIO{}-algorithm which produced additional degrees of freedom. Randomization of these additional degrees of freedom from iteration to iteration was exploited to prevent stagnation in local minima, especially in traps or tunnels. In particular, this was demonstrated for two pure phase objects with smooth phase variation (with different behavior with respect to the traditional \HIOER-algorithm) and a purely real object with strong amplitude variations over short length scales. The improved algorithm is far less sensitive to the specific choice of the internal parameters $N_{\mathrm{\HIO}}$ and $N_{\mathrm{\ER}}$ than the traditional \HIOER{}-algorithm. Moreover, the \HIOEROR{}-algorithm remains almost insensitive to the particular value of the internal parameter $\beta$ in the range $\beta \in [0.5, 1.0]$. Furthermore, the the good scaling properties in terms of computational time and memory consumption of the \HIOER{}-algorithm are fully maintained in the \HIOEROR{}-algorithm.  In conclusion, a reconstruction based on the \HIOEROR{}-algorithm exhibits significantly enhanced stability and success probability in comparison with the traditional and commonly used \HIOER{}-algorithm. 


\begin{thebibliography}{10}
\newcommand{\enquote}[1]{``#1''}

\bibitem{Veen2004}
F.~v.~d. Veen and F.~Pfeiffer, \enquote{Coherent x-ray scattering,} J. Phys.:
  Condens. Matter \textbf{16}, 5003--5030 (2004).

\bibitem{Millane1990}
R.~Millane, \enquote{Phase retrieval in crystallography and optics,} J. Opt.
  Soc. Am. A \textbf{7}, 394--411 (1990).

\bibitem{Pietsch2004}
U.~Pietsch, V.~Holy, and T.~Baumbach, \emph{High-Resolution X-ray Scattering
  From Thin Films to Lateral Nanostructures} (Springer, New York, 2004).

\bibitem{Afanasiev2004}
G.~N. Afanasiev, \emph{Vavilov-Cherenkov and Synchrotron Radiation: Foundations
  and Applications} (Springer, Netherlands, 2004).

\bibitem{Robinson2009}
I.~Robinson and R.~Harder, \enquote{Coherent x-ray diffraction imaging of
  strain at the nanoscale,} Nat Mater \textbf{8}, 291--298 (2009).

\bibitem{Miao1999}
J.~Miao, P.~Charalambous, J.~Kirz, and D.~Sayre, \enquote{Extending the
  methodology of x-ray crystallography to allow imaging of micrometre-sized
  non-crystalline specimens,} Nature \textbf{400}, 342--344 (1999).

\bibitem{Pfeifer2006}
M.~A. Pfeifer, G.~J. Williams, I.~A. Vartanyants, R.~Harder, and I.~K.
  Robinson, \enquote{Three-dimensional mapping of a deformation field inside a
  nanocrystal,} Nature \textbf{442}, 63--66 (2006).

\bibitem{Biermanns2009}
A.~Biermanns, A.~Davydok, H.~Paetzelt, A.~Diaz, V.~Gottschalch, T.~H. Metzger,
  and U.~Pietsch, \enquote{{Individual GaAs nanorods imaged by coherent X-ray
  diffraction},} Journal of Synchrotron Radiation \textbf{16}, 796--802 (2009).

\bibitem{Minkevich07}
A.~A. Minkevich, M.~Gailhanou, J.-S. Micha, B.~Charlet, V.~Chamard, and
  O.~Thomas, \enquote{Inversion of the diffraction pattern from an
  inhomogeneously strained crystal using an iterative algorithm,} Phys. Rev. B
  \textbf{76}, 104106 (2007).

\bibitem{Minkevich11EPL}
A.~A. Minkevich, E.~Fohtung, T.~Slobodskyy, M.~Riotte, D.~Grigoriev,
  T.~Metzger, A.~C. Irvine, V.~Novák, V.~Holý, and T.~Baumbach,
  \enquote{Strain field in ({G}a,{M}n){A}s/{G}a{A}s periodic wires revealed by
  coherent x-ray diffraction,} Europhysics Letters \textbf{94}, 66001 (2011).

\bibitem{Minkevich11PRB}
A.~A. Minkevich, E.~Fohtung, T.~Slobodskyy, M.~Riotte, D.~Grigoriev,
  M.~Schmidbauer, A.~C. Irvine, V.~Nov\'ak, V.~Hol\'y, and T.~Baumbach,
  \enquote{Selective coherent x-ray diffractive imaging of displacement fields
  in ({G}a,{M}n){A}s/{G}a{A}s periodic wires,} Phys. Rev. B \textbf{84}, 054113
  (2011).

\bibitem{Fienup1982}
J.~Fienup, \enquote{Phase retrieval algorithms: a comparison,} Appl. Opt.
  \textbf{21}, 2758--2769 (1982).

\bibitem{Fienup1986a}
J.~Fienup, \enquote{Reconstruction of a complex-valued object from the modulus
  of its fourier transform using a support constraint,} J. Opt. Soc. Am. A
  \textbf{4}, 118--123 (1986).

\bibitem{Levi84}
A.~Levi and H.~Stark, \enquote{Image restoration by the method of generalized
  projections with application to restoration from magnitude,} J. Opt. Soc. Am.
  A \textbf{1}, 932--943 (1984).

\bibitem{Youla1982}
D.~C. Youla and H.~Webb, \enquote{Image restoration by the method of convex
  projections: Part 1 -- theory,} Medical Imaging, IEEE Transactions on
  \textbf{1}, 81 --94 (1982).

\bibitem{PhysRevB.78.174110}
A.~A. Minkevich, T.~Baumbach, M.~Gailhanou, and O.~Thomas,
  \enquote{Applicability of an iterative inversion algorithm to the diffraction
  patterns from inhomogeneously strained crystals,} Phys. Rev. B \textbf{78},
  174110 (2008).

\bibitem{Millane1996}
R.~P. Millane, \enquote{Multidimensional phase problems,} J. Opt. Soc. Am. A
  \textbf{13}, 725--734 (1996).

\bibitem{Seldin1990}
J.~H. Seldin and J.~R. Fienup, \enquote{Numerical investigation of the
  uniqueness of phase retrieval,} J. Opt. Soc. Am. A \textbf{7}, 412--427
  (1990).

\bibitem{Bates1982}
R.~H.~T. Bates, \enquote{Fourier phase problems are uniquely solvable in more
  than one dimension,} Optik (Stuttgart) \textbf{61}, 247--262 (1982).

\bibitem{Auslander1989}
L.~Auslander and F.~A. Grunbaum, \enquote{The fourier transform and the
  discrete fourier transform,} Inverse Problems \textbf{5}, 149 (1989).

\bibitem{Elser2008}
V.~Elser and R.~P. Millane, \enquote{{Reconstruction of an object from its
  symmetry-averaged diffraction pattern},} Acta Crystallographica Section A
  \textbf{64}, 273--279 (2008).

\bibitem{Miao1998}
J.~Miao, D.~Sayre, and H.~N. Chapman, \enquote{Phase retrieval from the
  magnitude of the fourier transforms of nonperiodic objects,} J. Opt. Soc. Am.
  A \textbf{15}, 1662--1669 (1998).

\bibitem{Fienup1986b}
J.~Fienup and C.~Wackermann, \enquote{Phase-retrieval stagnation problems and
  solutions,} J. Opt. Soc. Am. A \textbf{3} (1986).

\bibitem{Bauschke02}
H.~H. Bauschke, P.~L. Combettes, and D.~R. Luke, \enquote{Phase retrieval,
  error reduction algorithm, and fienup variants: a view from convex
  optimization,} J. Opt. Soc. Am. A \textbf{19}, 1334--1345 (2002).

\bibitem{Takajo98}
H.~Takajo, T.~Takahashi, R.~Ueda, and M.~Taninaka, \enquote{Study on the
  convergence property of the hybrid input--output algorithm used for phase
  retrieval,} J. Opt. Soc. Am. A \textbf{15}, 2849--2861 (1998).

\bibitem{Takajo99}
H.~Takajo, T.~Takahashi, and T.~Shizuma, \enquote{Further study on the
  convergence property of the hybrid input--output algorithm used for phase
  retrieval,} J. Opt. Soc. Am. A \textbf{16}, 2163--2168 (1999).

\bibitem{Elser03}
V.~Elser, \enquote{Phase retrieval by iterated projections,} J. Opt. Soc. Am. A
  \textbf{20}, 40--55 (2003).

\bibitem{Marchesini2007Rev}
S.~Marchesini, \enquote{Invited article: A unified evaluation of iterative
  projection algorithms for phase retrieval,}  \textbf{78}, 011301 (2007).

\bibitem{Marchesini07}
S.~Marchesini, \enquote{Phase retrieval and saddle-point optimization,} J. Opt.
  Soc. Am. A \textbf{24}, 3289--3296 (2007).

\end{thebibliography}
\end{document}